\documentclass[sigplan,10pt]{acmart}
\usepackage{graphicx} 
\usepackage[utf8]{inputenc} 
\usepackage[T1]{fontenc}    
\usepackage{hyperref}       
\usepackage{url}            
\usepackage{booktabs}       
\usepackage{amsfonts}       
\usepackage{nicefrac}       
\usepackage{microtype}      
\usepackage{xcolor}         
\usepackage{svg}
\usepackage{subcaption}
\usepackage{multirow}
\usepackage{makecell}
\usepackage[ruled,vlined,linesnumbered]{algorithm2e}
\usepackage{tikz}
\usetikzlibrary{patterns, positioning}
\usepackage{enumitem}
\usepackage{soul}
\usepackage{dblfloatfix}
\usepackage[normalem]{ulem}

\newcommand{\name}{{DualScale}\xspace}
\newcommand{\namebaseline}{{PlaceOnly}\xspace}
\newcommand{\distserve}{{DistServe}\xspace}

\SetKw{KwBreak}{break}

\title{\name: Energy-Efficient Disaggregated LLM Serving via Phase-Aware Placement and DVFS}

\author{Omar Basit$^*$, Yunzhao Liu$^*$, Z. Jonny Kong, Y. Charlie Hu}
\thanks{* These authors contributed equally to this work.}
\affiliation{%
  \institution{Purdue University}
  \country{}
}

\date{February 2026}
\begin{document}
\renewcommand\footnotetextcopyrightpermission[1]{}


\newcommand{\ie}{i.e.,\xspace}
\newcommand{\eg}{e.g.,\xspace}
\newcommand{\etal}{et al.\xspace}
\newcommand{\parb}[1]{\vspace{1pt}\noindent{\bfseries #1.}}
\newcommand{\parbi}[1]{\vspace{1pt}\noindent{\bfseries\itshape #1.}}
\newcommand*\circled[1]{\tikz[baseline=(char.base)]{
            \node[shape=circle,draw,inner sep=2pt] (char) {#1};}}
\newcommand{\jonny}[1]{{\color{magenta} {(Jonny: #1)}}}
\newcommand{\omar}[1]{{\color{orange} {(Omar: #1)}}}
\newcommand{\yunzhao}[1]{{\color{magenta} {(Yunzhao: #1)}}}
\renewcommand{\comment}[1]{{\color{blue} {#1}}}
\newcommand{\cut}[1]{{}}

\def\Snospace~{\S}
\renewcommand{\sectionautorefname}{\Snospace}
\renewcommand{\subsectionautorefname}{\Snospace}
\renewcommand{\subsubsectionautorefname}{\Snospace}
\renewcommand*{\algorithmautorefname}{Algorithm}

\settopmatter{printfolios=true,printacmref=false}

\pagestyle{plain}
\setlength{\emergencystretch}{2em}

\begin{abstract}
Prefill/decode disaggregation is increasingly adopted in LLM serving
to improve the latency--throughput tradeoff and meet strict TTFT and
TPOT SLOs. However, LLM inference remains energy-hungry: autoscaling
alone is too coarse-grained to track fast workload fluctuations, and
applying fine-grained DVFS under disaggregation is complicated by phase-asymmetric
dynamics and coupling between provisioning and frequency control.

We present \name, a two-tier energy optimization framework for
disaggregated LLM serving. \name jointly optimizes placement and DVFS
across prefill and decode using predictive latency and power
models. At coarse timescales, \name computes phase-aware placement and
baseline frequencies that minimize energy while satisfying SLO
constraints. At fine timescales, \name dynamically adapts GPU
frequency per iteration using stage-specific control: model predictive
control (MPC) for prefill to account for queue evolution and future
TTFT impact, and lightweight slack-aware adaptation for decode to
exploit its smoother, memory-bound dynamics. This hierarchical design
enables coordinated control across timescales while preserving strict
serving SLOs.

Evaluation on a 16$\times$H100 cluster serving Llama~3.3~70B with
production-style traces shows that \name meets TTFT/TPOT SLOs while reducing
energy by up to 39\% in prefill and 48\% in decode relative to DistServe.
\end{abstract}

\maketitle

\section{Introduction}

Recent frontier models such as GPT-5, Claude, Kimi K2, and
DeepSeek-V3~\cite{openai-gpt5-2025,anthropic-models-overview-2025,kimi-k2-2025,liu2024deepseek}
now power a wide range of user-facing services, including chatbots, coding
assistants, and enterprise agents.
However, these models impose substantial compute and memory demands per
request, making it fundamentally challenging to simultaneously meet strict
latency SLOs and sustain high throughput.
In these interactive settings, user experience depends on strict latency
service-level objectives. In particular, users expect low \textit{Time To
First Token} (TTFT) and stable \textit{Time Per Output Token} (TPOT).
At the same time, providers must maintain high throughput to keep serving costs
under control.

To address this tension between strict latency SLO targets and high throughput,
recent LLM serving systems increasingly adopt prefill/decode disaggregation,
where prompt processing and autoregressive generation are provisioned
separately.
Recent systems such as DistServe~\cite{distserve:osdi} and
Splitwise~\cite{splitwise:isca2024} show that this design improves the
latency-throughput tradeoff and helps meet strict TTFT and TPOT SLOs.
The same direction is visible in industry: DeepSeek-V3 ~\cite{liu2024deepseek}
and serving stacks such as NVIDIA Dynamo and
vLLM~\cite{nvidia-dynamo-disagg-2025,vllm-disagg-2025} all support
disaggregated serving.

Despite these efficiency gains, the energy footprint of LLM inference remains
substantial. 
Recent production-oriented analysis estimates a median of 0.34 Wh per query for
frontier-scale models on H100-class inference systems, and up to 4.32 Wh per
query under test-time scaling workloads~\cite{microsoft-inference-energy-2025}.
At deployment scale, this corresponds to roughly 0.8--1.8 GWh/day for a system
serving 1B queries per day~\cite{microsoft-inference-energy-2025}.
At the serving-system level, recent characterization work reports that
inference clusters often operate with significant power overhead under
time-varying workloads~\cite{patel2024characterizing,energy_benchmark:hotcarbon2025}.

Many serving systems reduce energy indirectly via autoscaling, by adapting
active resources to time-varying workload
demand~\cite{inferline:osdi2020,serverlessllm:osdi2024}.
However, these mechanisms operate at coarse timescales because reconfiguration
incurs non-trivial overheads, such as model weight loading, and therefore
cannot swiftly follow fine-grained workload
fluctuations ~\cite{nvidiacoldstart}.
Therefore, they need to over-provision according to the peak load
over  the period
of time of each configuration, leading to inefficiencies during non-peak times.

%
Modern GPUs support Dynamic Voltage and Frequency Scaling (DVFS), which allows
the system to adjust SM clock frequency (and corresponding voltage states) at
fine time scales, \eg on the order of tens of milliseconds.
DVFS has been successfully applied in LLM training to reduce energy consumption by up to 30\% with negligible performance impact~\cite{you2023zeus,chung2024reducing}.
However, applying DVFS to online inference presents fundamentally new challenges.
First, prefill and decode have distinct bottlenecks and latency sensitivities, and 
must continuously satisfy their respective SLOs (TTFT for prefill and TPOT for decode)
as the workload fluctuates over time;
second, under disaggregation, energy efficiency depends jointly on cluster placement decisions and DVFS settings across both phases.
%

Several recent LLM serving systems hav exploited DVFS to reduce the
energy drain, including DynamoLLM~\cite{dynamollm:hpca2024} and
throttLL'eM~\cite{throttllem:hpca25}.  However, these work focus on
non-disaggregated architectures, and therefore do not address the
joint optimization required in disaggregated serving.

Motivated by this gap, in this paper, we study minimizing energy
consumption in prefill/decode-disaggregated LLM serving.
Unlike prior non-disaggregated energy-efficient serving settings, disaggregation
introduces additional coupling across stages and timescales, leading to several
new challenges:

\begin{itemize}[leftmargin=1em,noitemsep,topsep=0pt]

\item \textit{How to pick cluster placement configurations under a large search space?}
In disaggregated serving, provisioning must jointly configure \emph{both} prefill
and decode pools. A placement must choose, for each phase, the number of
instances, parallelism configurations, baseline frequencies, and routing weights.
The cross-product of these phase-specific choices yields a large combinatorial
search space. Moreover, the two phases are tightly coupled: under-provisioning
prefill degrades TTFT, while under-provisioning decode creates post-prefill
backlogs and inflates TPOT, even if the other phase is over-provisioned.
Because input/output length distributions evolve over time and shift load
pressure between phases over minutes, the energy-optimal placement must be
recomputed efficiently to track workload changes.

\item \textit{How should DVFS control adapt to phase-asymmetric dynamics?}
Prefill and decode have fundamentally different bottlenecks, memory pressure, and
temporal behavior, which makes uniform DVFS ineffective. Prefill is typically
compute-bound and highly frequency-sensitive, whereas decode is often
memory-bandwidth-bound and retains large KV caches across many iterations,
creating sustained memory pressure. Prefill load is arrival-driven and bursty,
while decode load evolves more smoothly as requests remain in decode for many
iterations. Effective DVFS therefore must be phase-aware and tailored to each
stage’s dynamics to preserve TTFT/TPOT SLOs.
\end{itemize}

To address these challenges, we design \name, a two-tier energy optimization framework for disaggregated LLM serving. \name is built on the key insight that energy-efficient serving requires coordinated control across both coarse-grained provisioning and fine-grained DVFS, while respecting the distinct dynamics of prefill and decode.

At Tier~1, \name computes phase-aware placement and baseline frequency configurations at coarse timescales using predictive latency and power models. Tier~1 identifies an energy-minimizing operating point that satisfies TTFT and TPOT SLOs under expected peak load, ensuring system feasibility.
At Tier~2, \name dynamically adjusts GPU frequency at iteration granularity to exploit short-term workload slack and correct prediction errors. \name employs stage-specific control strategies tailored to phase dynamics. For prefill, where frequency decisions affect queue evolution and future TTFT, \name uses model predictive control (MPC) to minimize energy while preserving latency headroom over a multi-batch horizon. For decode, which exhibits smoother and predominantly memory-bound behavior, \name applies lightweight per-iteration frequency adaptation to safely harvest slack.

Both tiers rely on data-driven iteration-level latency and power models trained offline, enabling accurate prediction of performance and energy across configurations. Together, this hierarchical design enables \name to jointly optimize placement and DVFS across timescales, substantially reducing energy while preserving strict serving SLOs.

We implement \name and evaluate it using Llama~3.3~70B on a 16$\times$H100
cluster.
We implement \name and evaluate it using Llama~3.3~70B on a 16$\times$H100 cluster. Both tiers contribute to the overall energy savings.
Relative to DistServe,
Tier~1 alone (PlaceOnly) reduces prefill energy by up to 29\% and decode energy by up to 45\%.
Tier~2 (\name) further improves energy efficiency via fine-grained DVFS, increasing the reduction to up to 39\% for prefill and 48\% for decode.

In summary, this paper makes the following contributions:
\begin{itemize}[leftmargin=1em,noitemsep,topsep=0pt]
    \item We present, to our knowledge, the first energy-efficient LLM
        inference system under prefill/decode disaggregation. We characterize
        how phase asymmetry and workload dynamics jointly shape the energy-SLO
        tradeoff.
    \item We design \name, a two-tier solution that combines phase-aware
        coarse-grained placement with fine-grained online DVFS. The design
        explicitly captures placement--DVFS coupling and uses DVFS as an online
        correction mechanism for workload-prediction error.
    \item We implement and evaluate \name on production-style traces and a real
        multi-GPU testbed. Results show substantial energy reductions over
        strong baselines while preserving strict TTFT/TPOT SLOs.
\end{itemize}

\section{Background}

\cut{
\subsection{LLM Serving}
Large Language Model (LLM) serving refers to the process of deploying trained models to execute inference requests. Unlike traditional deep learning inference, LLM serving is characterized by extreme computational and memory requirements, primarily due to the immense size of model parameters and the transient state required during generation.

\subsubsection{LLM Serving Phases}
The execution of an LLM inference request is distinctively divided into two sequential phases, often referred to as ``Prefill'' and ``Decode.''

\begin{itemize}
    \item \textbf{Prefill Phase:} Upon receiving a prompt, the system processes all input tokens in parallel to calculate the initial hidden states. Because the input tokens are available simultaneously, this phase allows for dense matrix operations, making it highly parallelizable and generally compute-bound. 
    
    \item \textbf{Decode Phase:} Following prefill, the model enters an autoregressive generation loop. It generates one token at a time, with each new token depending on the entire history of previous tokens. This sequential dependency prevents parallelization across the sequence dimension. Consequently, this phase is typically memory-bandwidth bound, as the system must additionally load the context data for every single token generated.
\end{itemize}

\subsubsection{Key-Value (KV) Caching}
To mitigate redundant computation during the decode phase, LLM serving systems utilize KV Caching. In the standard Transformer self-attention mechanism, the attention output for a query token depends on the Key ($K$) and Value ($V$) vectors of all preceding tokens.

Without caching, generating the $i$-th token would require recomputing the $K$ and $V$ projections for all tokens from $1$ to $i-1$, resulting in quadratic computational complexity. The KV cache addresses this by storing the $K$ and $V$ vectors in GPU memory as they are computed, which can then be recalled for successive token generation in the decode phase.

\subsubsection{Performance Metrics}
To evaluate the efficiency of a text-based LLM serving system, distinct metrics are utilized to capture user experience.

\begin{itemize}
    \item \textbf{Time to First Token (TTFT):} This measures the latency from the moment a request arrives until the system outputs the first token. It depends on the speed of the prefill phase and the request queuing time. A low TTFT is critical for interactive applications.

    \item \textbf{Time Per Output Token (TPOT):} This measures the average time elapsed between the generation of consecutive tokens for a request during the decode phase. Smooth user experiences require the TPOT to be lower than the user's reading speed.

    
\end{itemize}
}

\subsection{LLM Online Serving Workload Dynamics}
\label{subsec:workload_dyanmics}

\begin{figure}[tp]
	\centering
    \includegraphics[width=0.8\columnwidth,trim=0 0 0 0]{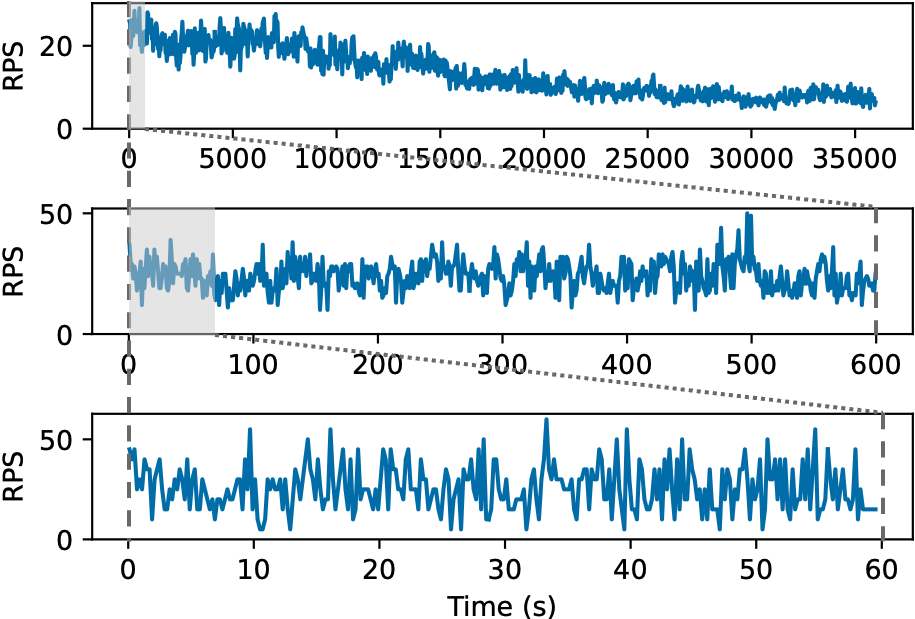}
    \vspace{-8pt}
    \caption{
        The RPS timelines of the Azure LLM inference
        trace~\cite{azure-public-dastaset} over 10 hours, 10 minutes, and 1
        minute. 
    }
    \label{fig:qps-timelines}
    \vspace{-4pt}
\end{figure}

\begin{figure}[tp]
	\centering
    \includegraphics[width=0.7\columnwidth,trim=0 0 0 0]{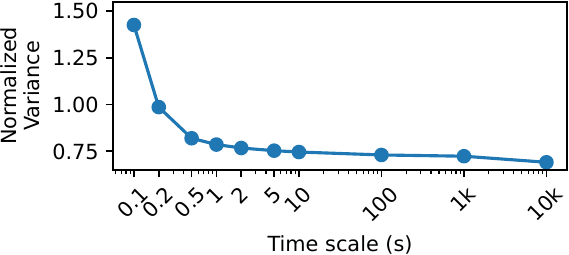}
    \vspace{-8pt}
    \caption{
        Variance-time plot of request-per-second (RPS) in the Azure LLM inference
        trace~\cite{azure-public-dastaset}. The trace exhibits notable
        fluctuation across both short and long timescales, with slightly
        greater variance observed at shorter timescales.
    }
    \label{fig:variance-time-plot}
    \vspace{-4pt}
\end{figure}

\begin{figure}[t]
    \centering
    \includegraphics[width=0.8\columnwidth]{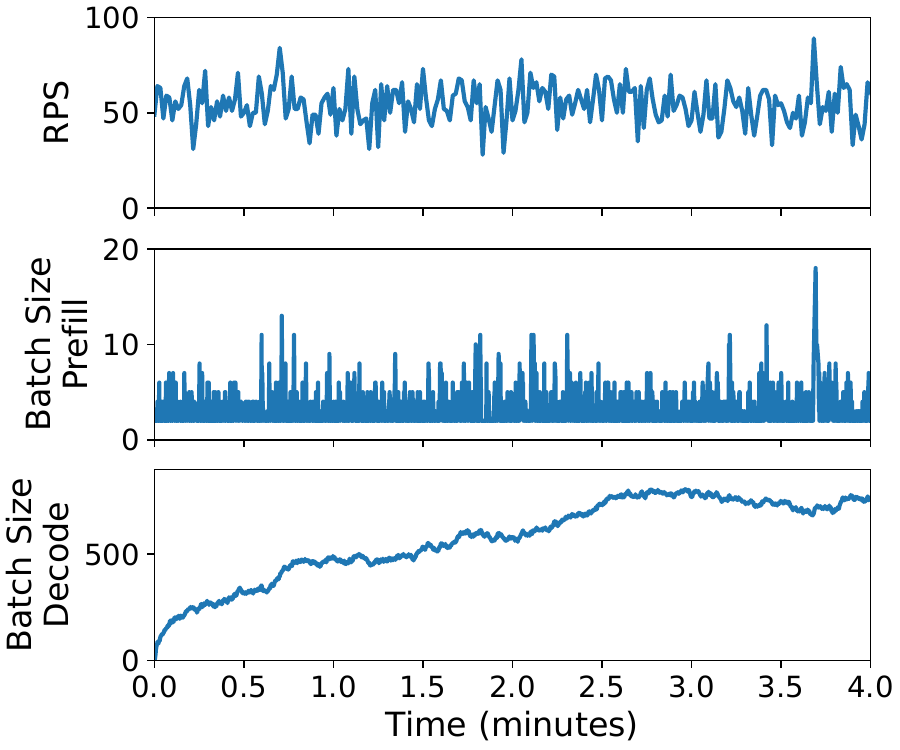}
    \vspace{-8pt}
    \caption{
        Number of running requests in prefill and decode instances plotted with
        workload in RPS.
    }
    \label{fig:rps_batch}
    \vspace{-4pt}
\end{figure}

Online user-facing LLM inference systems receive requests with workload that
dynamically changes over time.
In \autoref{fig:qps-timelines}, we show the RPS timeline from the Azure LLM
inference trace~\cite{azure-public-dastaset} across multiple time scales. The trace
shows that the workload exhibits burstiness consistently across different
timescales -- from hours to minutes to seconds.

To quantify this workload fluctuation, we show the \textit{normalized
variance-time plot}~\cite{garrett1994analysis,leland2002self} of the full
trace, which spans 7 days. 
To calculate normalized variance-time, we first divide the trace into
non-overlapping windows with sizes ranging from 0.1 to 10000 seconds.
For each window, we compute the request per second (RPS), then calculate the
mean and variance of these RPS values across all windows. We plot the
normalized variance, defined as the ratio of variance to mean, as a function of
window size, as shown in \autoref{fig:variance-time-plot}.

We observe significant workload variability across multiple time scales. At the
hourly level (1,000–10,000s), the normalized variance stabilizes around 0.7,
reflecting diurnal and daily load cycles. 
At the minute scale (100–1,000 s), the variance increases further. 
At the second scale (down to 0.1 s), the trace exhibits sharp RPS swings, with normalized variance peaking at 1.4.
This shows that online LLM serving workloads vary significantly at both
long and short timescales, 
and LLM inference systems must be designed to tolerate this variability.

\subsection{Energy Efficient LLM Serving}

To reduce the energy consumption of LLM inference serving, existing systems
typically rely on two complementary mechanisms that operate at different
time scales.
First, many existing systems employ autoscaling to address long-term workload
variations~\cite{inferline:socc2020,deepspeed:sc2022,aladdin:2024arxiv}.
By dynamically adjusting the number of active serving instances based on
historic request rates, autoscaling reduces idle resources during low-load
periods and hence improves overall utilization.
However, autoscaling is limited by high reconfiguration overheads in LLM
serving, including model loading and request draining, which typically take
several minutes ~\cite{nvidiacoldstart}.
As a result, it cannot respond to fine-grained workload fluctuations.

Second, Dynamic Voltage and Frequency Scaling (DVFS) reduces energy consumption
by adjusting GPU operating frequency.
Unlike autoscaling, DVFS can operate at both long and short time scales,
ranging from coarse-grained power provisioning to fine-grained adaptation under
bursty workloads.
Prior works have shown that GPUs often run at unnecessarily high frequencies
during memory-bound or lightly loaded phases, wasting
energy~\cite{dynamollm:hpca2024,stojkovic2024towards,throttllem:hpca25}.
Accordingly, these works reduce power consumption by lowering GPU operating
frequency when possible.

Recent systems integrate these ideas with inference-specific optimizations.
For example, DynamoLLM~\cite{dynamollm:hpca2024} improves energy efficiency by
partitioning GPUs into pools based on input and output lengths, and selects the
optimal parallelism and frequency configurations within each pool.
throttLL'eM~\cite{throttllem:hpca25} applies predictive DVFS to throttle GPU
frequency during inference execution, reducing energy while maintaining latency
SLOs.

\cut{
\subsubsection{Coarse-grained: Minute-Level Variation}
As shown in \autoref{fig:variance-time-plot}, workload variations at the
minute-to-hour scale are relatively stable, allowing systems to predict demand
and adjust resources accordingly.
For systems that do not optimize for power, this slow-moving variability is
usually managable, as the high predictability of the large-timescale average
workload simplifies capacity planning.
This predictability provides a window to calculate and deploy an optimal
placement—defined here as a configuration, which includes the number of GPUs,
their parallelism, their load distribution and frequency, such that energy
consumption is minimized while meeting the SLOs.
 

However, a challenge arises when we try to search for this optimal placement:
the large search space.
To find the most energy-efficient placement, the system must consider GPU
frequencies as a decision variable.
Since lower frequencies can significantly reduce power consumption during
periods of moderate load, they must be included in the placement logic.
Including these frequency settings drastically increases the complexity of the
search space compared to existing autoscaling systems that do not adjust the
frequency.

\subsubsection{Fine-grained: Iteration-Level Variation} \label{sec:challenge-fine}



The high degree of variation in request arrivals at second and sub-second
scales, illustrated in \autoref{fig:variance-time-plot}, creates significant
challenges for autoscaling-based systems.
Because reconfigurations cannot be executed frequently due to time cost, any
placement strategy targeting the minute scale must provision for the peak
workload to ensure SLO compliance.
Consequently, the system remains under-utilized during the frequent ``dips'' in
traffic that occur between these peaks. During these intervals, the static
configuration consumes excessive power relative to the actual demand.

DVFS is a potential mechanism to capture these missed energy-saving
opportunities.
%
%
By adjusting the GPU clock frequency in a fine-grained manner, DVFS allows the
system to throttle power consumption in real-time whenever the workload falls
below the provisioned peak.
Applying DVFS on a disaggregated system requires us to separately determine the
frequencies of prefill and decode stages based on their characteristics. In
\autoref{fig:rps_batch}, we plot a synthetic bursty workload that we ran on a
disaggregated system, and the resulting behaviors of the prefill and decode
instances in terms of the requests contained in each running batch.

\textbf{Prefill:}
In prefill instances, all requests will finish in a single iteration, or
several iterations if chunked prefill~\cite{agrawal2024taming} is enabled and
when the request's prompt length is above the chunking budget.
%
Consequently, each request stays in the prefill stage for a short duration,
typically a single iteration, before moving on to the decode stage.
This means that the bursty request pattern is retained at the sub-minute range
and is translated into a bursty pattern of GPU compute usage, ranging from idle
to complete usage. 
This effect is visible in \autoref{fig:rps_batch}, where the batch size varies
from just 1 when only a single request arrives, and when bursts arrive, we see
them as spikes in the batch size.
%
%
Such variance is where a coarse-grained approach, such as autoscaling, will
struggle in terms of energy efficiency, and we need power saving mechanisms
such as DVFS that can operate in a fine grained manner.

\textbf{Decode:}
In decode instances, since tokens are generated one by one, requests stay in
the decode stage for large number of iterations --- as high as their output
length -- which can be up to thousands.
This means the bursty arrival of requests gets smoothed over, \ie the compute
on the decode instance has ``momentum'' and changes slowly. This is visible in
the decode batch size in \autoref{fig:rps_batch}, where the spikes and valleys
of the workload is much smoother than the prefill batch size.

%
However, the reduced workload fluctuation on decode instances does not mean
that coarse-grained autoscaling systems is sufficient.
In the previous section, we briefly discussed using predicted workload to get
an energy-optimal configuration. If this prediction is off, we may get a
configuration that is either able to maintain SLO but is not energy efficient,
or that is energy efficient but violates the SLO.
So, although decode stage workload fluctuation is less significant than that of
the prefill stage, coarse-grained autoscaling systems still suffer from
workload mis-prediction.
%
}

\subsection{Prefill/Decode Disaggregation}
In many existing LLM serving systems
\cite{orca:osdi2022,deepspeed:sc2022,kwon2023efficient,agrawal2024taming}, the
prefill and decode phases are collocated on the same GPU, interleaving the
compute-bound prefill of new requests with the memory-bound decode of ongoing
requests, which is known as continuous batching~\cite{orca:osdi2022}.
%
%
However, such systems suffer from significant inter-phase interference.
Prefill workloads are bursty and compute-intensive; when a large prefill
request is scheduled, it can delay the execution of ongoing decode steps,
leading to spikes in inter-token latency.
This effect, effectively a form of head-of-line blocking, makes it difficult to
simultaneously satisfy the service level objectives (SLOs) for prefill and
decode, namely time-to-first-token (TTFT) and time-per-output-token (TPOT).
Prior work has proposed mitigations such as chunked
prefill~\cite{agrawal2024taming}, which limits the number of tokens processed
per iteration by splitting prefills into smaller chunks.
However, selecting an appropriate chunk size that balances TTFT and TPOT
remains challenging and is workload-dependent.

To address this problem, recent systems adopt \textbf{Prefill-Decode (P/D)
Disaggregation}~\cite{distserve:osdi,splitwise:isca2024}, which partitions the GPUs 
into two specialized pools of instances:
1) Prefill instances, which process prompts and compute the initial key–value
(KV) cache.
2) Decode instances, which are dedicated to autoregressive token generation.
Once a request finishes prefill on a prefill instance, it is then forwarded to
a decode instance along with its KV cache.
In this disaggregated setup, prefill and decode no longer interfere with each
other, and the two phases can both maintain low and stable latencies.
%
%
This separation also enables independent scaling of the two stages, allowing
resources to be allocated according to the prefill-to-decode workload ratio.
%
%
For example, DistServe reports up to a $7.4\times$ improvement in goodput over
collocated baselines, while meeting TTFT and TPOT SLOs.


Beyond improved goodput, disaggregated architectures provide greater
flexibility for system reconfiguration under the presence of fluctuating
workloads.
Although dynamic autoscaling is commonly used in online services to improve
utilization~\cite{lorido2014review}, its effectiveness in LLM serving is
limited by high reconfiguration costs.
Reconfiguring a serving instance requires loading large model parameters into
GPU memory and draining in-flight requests, incurring substantial
latency~\cite{serverlessllm:osdi2024}.
As a result, reconfiguration is typically coarse-grained.
Disaggregation mitigates this limitation by allowing prefill and decode pools
to scale independently, enabling targeted adjustments to phase-specific demand
and reducing the frequency and impact of disruptive reconfigurations compared
to collocated systems.


\subsection{The Energy Efficient P/D Disaggregation Serving Problem}

Given the benefits of P/D disaggregation and the growing importance of energy
efficiency, in this paper we study the following problem:
\textit{How to minimize the energy consumption of a two-level P/D-disaggregated LLM
inference serving system while meeting strict TTFT and TPOT SLOs?}

\cut{
\subsection{DVFS as a Mechanism for Power Saving} While the exorbitant energy
cost of \textit{training} LLMs has historically garnered the most attention,
the cumulative energy footprint of \textit{inference} is rapidly becoming the
dominant factor in the lifecycle carbon emissions of these models. Recent
analysis suggests that as LLMs are deployed at scale (e.g., ChatGPT), the
aggregate inference compute demand (and associated carbon footprint) outgrows
the one-time cost of training~\cite{energy_benchmark:hotcarbon2025}.

In traditional online serving environments, energy efficiency is often
approached through autoscaling, which indirectly reduces power by maximizing
hardware utilization and shutting down idle instances during low-traffic
periods. However, while autoscaling manages resource allocation at the node
level, Dynamic Voltage and Frequency Scaling (DVFS) provides a mechanism for
managing the power profile of the underlying hardware. DVFS operates by
dynamically modulating the operating frequency ($f$) and supply voltage ($V$)
of the processor’s cores and memory subsystems in response to workload
characteristics. While long established in mobile and power-constrained
systems, DVFS has gained significant traction in the server space as inference
systems increasingly rely on power-hungry GPUs.

In the context of Deep Neural Networks (DNNs) and LLMs, DVFS has been shown to
successfully minimize energy waste. Frameworks such as Zeus~\cite{you2023zeus}
and Perseus~\cite{chung2024reducing} demonstrate that GPUs, by default, run at
maximum frequency even when the workload is memory-bound or when stragglers in
a distributed pipeline create slack. Perseus applies DVFS in LLM training by
identifying non-critical paths or memory-bound phases, and reduces frequency to
save energy without violating Service Level Objectives (SLOs) or increasing
end-to-end iteration time. Zeus identifies a Pareto frontier between energy
consumption and latency in DNNs, allowing operators to trade marginal
performance losses for significant energy gains using frequency scaling.
}

\section{Challenges}
\label{sec:challenges}

\cut{
Generally, inference systems facing fluctuating online workloads rely on
autoscalers to maximize resource utilization. But its effectiveness is capped
by the workload's predictability. If workload variations outpace the
autoscaler's ability to react, utilization drops, and energy efficiency
suffers, rendering placement configurations insufficient on their own.

Consequently, as noted in prior work
\cite{dynamollm:hpca2024,throttllem:hpca25,slo-aware:2024letter},
energy-optimized serving must operate across two distinct dimensions. The first
is coarse-grained placement where we tune the model placements, \ie parallelism
strategies and the number of GPUs, to find one that is more energy-efficient
for the current coarse-grain load.
The second is to use DVFS on a fine-grained time scale to tune the frequency of
the serving instances based on the load, \ie reduce frequency when the load on
an instance is low and vice versa.

Workload variability dictates which optimization strategy is effective at a
specific time scale. We characterize the workload using two primary metrics:
average Requests per Second (RPS) and the request arrival pattern. While RPS
informs coarse-grained capacity planning, the arrival pattern is the critical
factor for fine-grained energy management, as we will discuss next.
}

By separating the prefill and decode phases onto different execution pools,
disaggregation improves SLO attainment but introduces new challenges for
energy-efficient serving.
In particular, disaggregation exposes pronounced {\it asymmetry} in workload
characteristics, resource requirements, and SLO sensitivity between the two
phases.
This asymmetry complicates both resource provisioning
(to exploit coarse-grained traffic variability)
and DVFS control (to exploit-grained traffic variability).

\subsection{Key Observation: Phase-specific Workload Characteristics}
\label{subsec:phase}

A key observation in LLM serving is that the prefill and decode phases exhibit
fundamentally different workload characteristics~\cite{agrawal2024taming}.
First, since prefill phase is typically compute-bound whereas decode phase
tends to be memory-bandwidth-bound, prefill and decode latencies have
different sensitivities to input length and GPU clock frequency.
For example, prior work~\cite{stojkovic2024towards} shows that prefill latency (\ie
TTFT) is more heavily affected by input length and GPU frequency, while
decode latency (\ie TPOT) is less sensitive to frequency and is
largely insensitive to input length on modern GPUs that have high memory
bandwidth.

The distribution of request input and output lengths further
affects resource demand and consequently how DVFS should be applied across
prefill and decode.
Workloads with long prompts and short
responses place more pressure on prefill, whereas workloads with short prompts
and long generations stress decode throughput and memory bandwidth.

\subsection{Key Challenges}

The above  phase-specific bottlenecks and workload variations 
present a unique set of challenges in exploiting resource provisioning and DVFS control
to optimize energy efficiency in a P/D disaggregated serving system.

\textbf{C1: Combinatorial search space of energy-efficient
  resource provisioning.}
At coarse timescales, resource provisioning periodically reconfigures the
cluster, \ie the placement of GPUs across serving instances, to minimize
cluster energy while meeting TTFT/TPOT SLOs for the next workload window.
In non-disaggregated systems, provisioning only needs to choose the number of
instances and their configurations (\eg tensor-parallel degree and operating
frequency).
{
Under P/D disaggregation, however, the system must provision
\emph{both} prefill and decode pools, and the two decisions are tightly coupled:
under-provisioning prefill degrades TTFT, while under-provisioning decode
creates post-prefill backlogs and inflates TPOT; over-provisioning either phase
wastes energy or simply shifts the bottleneck to the other phase. Consequently,
provisioning must jointly choose phase-specific instance counts and
configurations, yielding a large combinatorial configuration space.

This problem is further complicated by workload variations. Because prefill and
decode have different resource bottlenecks and the input/output length
distribution evolves over time, the dominant load can shift between phases,
changing the energy-optimal provisioning. As we show in \autoref{subsubsec:production_workload},
these shifts can occur within minutes.
Therefore, {\it provisioning must search a large space and do so fast
enough to continuously track workload shifts}.
}
%

\textbf{C2: Phase-specific workload characteristics
complicates DVFS control.}
Exploiting fine-grained workload variability using DVFS is particularly
challenging under P/D disaggregation due to fundamental differences between
prefill and decode.
First, as discussed in \autoref{subsec:phase}, the two phases have distinct
performance bottlenecks and frequency sensitivities: prefill is typically
compute-bound and highly sensitive to GPU frequency, whereas decode is often
memory-bandwidth-bound and less responsive to frequency scaling.
Second, the two phases differ significantly in GPU memory pressure.
{While prefill instances process prompts without retaining large KV caches,
decode instances must maintain per-request KV states across many
iterations,} creating sustained GPU memory pressure and increasing the risk of
out-of-memory (OOM) conditions. DVFS decisions must therefore account for both
compute performance and memory feasibility.

Third, prefill and decode exhibit distinct temporal dynamics. Decode
load evolves smoothly because each request remains in the decode stage for many
iterations—one per generated token—often up to thousands. In contrast, prefill
load is arrival-driven and can appear as short-lived but intense spikes, since
a request typically completes prefill in a single iteration (or a small number
of iterations with chunked prefill~\cite{agrawal2024taming}).
\autoref{fig:rps_batch} illustrates this effect by plotting prefill
and decode batch sizes as proxies for phase-specific load
intensity 
higher variability than decode (bottom plot).

These differences in bottlenecks, memory pressure, and
temporal dynamics make it difficult to apply a uniform DVFS policy across both
phases. Instead, {\it effective DVFS must be phase-aware and responsive to the
distinct characteristics of prefill and decode.}

\section{Design}

\subsection{Design Principles}

Our objective is to minimize the energy consumption of a P/D-disaggregated LLM
serving system while satisfying strict TTFT and TPOT SLOs under dynamically
varying workloads.
As discussed in \autoref{sec:challenges}, this problem is challenging due to
(i) the asymmetric performance characteristics of prefill and decode, and
(ii) time-varying workload intensity and input/output length distributions.

\textbf{Key insight: hierarchical control with coordinated feasibility and adaptation.}
\name is built on the observation that energy-efficient serving
requires separating feasibility enforcement from fine-grained
energy optimization, while coordinating both through predictive
models. Placement decisions determine the feasible operating
region by provisioning sufficient compute, memory, and latency
headroom. Within this feasible region, fine-grained DVFS can
safely exploit transient workload slack to reduce energy.
To realize this insight, \name adopts a {\it two-tier control architecture}
operating at distinct timescales.

\textbf{Tier 1: Phase-aware coarse-grained provisioning.}
Every provisioning window (e.g., 5 minutes), Tier 1 determines
(i) the number of prefill and decode instances,
(ii) the tensor parallel (TP) degree of each instance,
(iii) a baseline operating frequency for each instance, and
(iv) request routing weights.
Tier 1 uses predictive latency and power models to identify an
energy-minimizing configuration that satisfies TTFT and TPOT
constraints under expected peak load. This ensures that all subsequent
DVFS actions operate within a provably feasible region.

\textbf{Tier 2: Phase-specific fine-grained DVFS control.}
At iteration granularity, Tier 2 dynamically adjusts GPU frequency
to exploit short-term workload slack while preserving SLOs.
Tier 2 never violates feasibility guarantees established by Tier 1.
Instead, it adapts frequency to workload dynamics, correcting
prediction errors and harvesting slack caused by burstiness and
load variation. Tier 2 applies stage-specific control strategies
that account for the distinct temporal and resource characteristics
of prefill and decode.

This hierarchical decomposition separates long-term capacity
provisioning from short-term energy adaptation, enabling scalable,
model-driven optimization across both timescales while preserving
strict SLO guarantees.

\subsection{System Architecture}

\begin{figure*}[t]
    \centering
    \includegraphics[width=\textwidth]{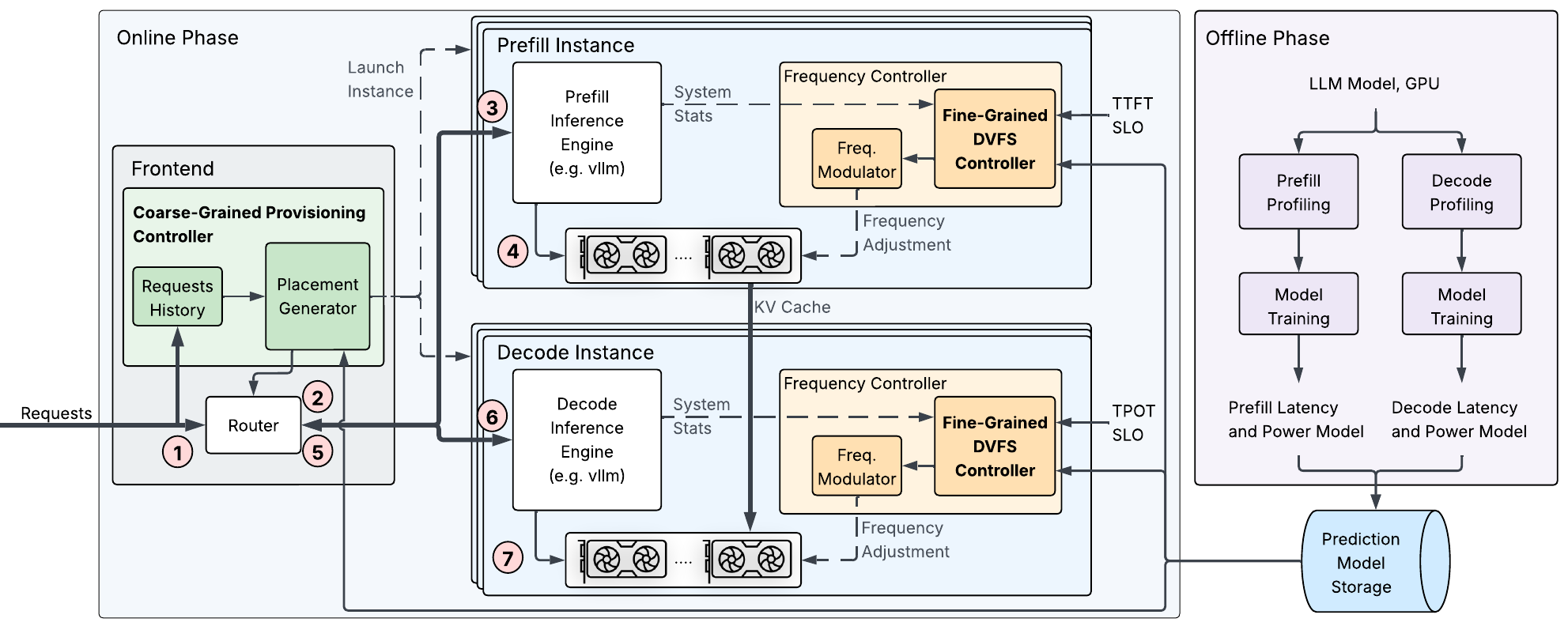}
    \vspace{-16pt}
    \caption{Architecture overview of \name.}
    \label{fig:overview}
    \vspace{-8pt}
\end{figure*}

Figure~4 illustrates the \name architecture, which consists of an
offline profiling phase and an online serving phase. The online phase
includes a \emph{data plane} (prefill and decode instances that execute
inference) and a \emph{control plane} (a coarse-grained provisioning controller
and a fine-grained DVFS controller).

\textbf{Offline profiling.}
\name trains latency and power models for each (LLM model, GPU type) pair by
profiling execution across different batch shapes, tensor-parallel (TP) degrees,
and GPU frequencies. These models capture iteration-level latency and energy
consumption and are used by both tiers at runtime.

\textbf{Online serving: request path and data plane.}
When a request arrives at the frontend, it is first sent to the router
(\textcircled{1}). The router selects a prefill instance based on the current
routing weights and forwards the request (\textcircled{2}). Inside the prefill
instance, the request enters a waiting queue (\textcircled{3}) and is later
scheduled for execution (\textcircled{4}). After prefill completes, the request
returns to the router (\textcircled{5}), which forwards it to a decode instance
according to the routing weights. The request then queues at the decode instance
(\textcircled{6}) and is scheduled for token generation (\textcircled{7}); during
decode, the instance retrieves the corresponding KV cache produced by prefill.

\textbf{Online serving: control plane.}
The coarse-grained provisioning controller periodically analyzes recent workload
history (recorded at request arrival in \textcircled{1}) and computes a placement
plan for the next provisioning window, including instance counts, TP degrees,
routing weights, and baseline frequencies. It launches or tears down prefill and
decode instances as needed and updates the router’s load-balancing weights.

The fine-grained DVFS controller continuously monitors runtime statistics from
each instance, including queue lengths and batch characteristics (collected
during \textcircled{3}--\textcircled{4} for prefill and \textcircled{6}--\textcircled{7}
for decode), and adjusts GPU frequency at iteration granularity via a lightweight
frequency modulator. The DVFS controller targets the TTFT SLO for prefill and the
TPOT SLO for decode while minimizing power consumption.

Both controllers rely on the offline-trained latency and power models to predict
SLO feasibility and energy under candidate configurations, enabling coordinated,
energy-efficient placement and DVFS decisions under dynamically varying workloads.

\subsection{Tier 1: Coarse-Grained Provisioning}
\label{subsec:tier1}

Tier~1 establishes an energy-minimizing operating point that guarantees
SLO feasibility under predicted peak load, defining a safe operating region
within which Tier~2 can dynamically adapt frequency.

{
\subsubsection{The Cluster Provisioning Problem}

Tier 1 reacts to longer-term workload shifts.
by periodically, \eg every 5 minutes,
determining a placement configuration for prefill and decode
that minimizes the total energy of the cluster while meeting TTFT/TPOT SLOs at a target goodput $R$,
\ie the predicted peak request rate during the next provision window.
{
 Similar to prior work such as DistServe~\cite{distserve:osdi},
 \name predicts the next-window workload, \eg request trace,
 using recent request history.
}

%
Specifically, we define a \textit{serving instance configuration} as a tuple:
\[
(type, \; TP\ degree, \; frequency)
\]
where $type \in \{prefill, decode\}$.
A \textit{placement configuration} for the cluster then
specifies (1) the number of instances of each serving instance configuration,
(2) the TP degree of each instance, (3) the baseline GPU frequency of each
instance, and (4) routing weights across instances.
}

In addition to SLO
feasibility, Tier~1 must ensure memory feasibility for both phases under the
predicted input/output length distribution.

\subsubsection{Placement Optimization via ILP}

{
Provisioning prefill and decode jointly creates a large combinatorial
search space due to the cross-product of instance types, TP degrees,
and frequencies. To efficiently find an optimal configuration under
resource and SLO constraints, \name formulates provisioning as an
integer linear programming (ILP) problem.

Given a total target goodput $R$, the ILP outputs the optimal placement that
minimizing energy under cluster capacity and SLO constraints, which includes:
\begin{itemize}
\item how many prefill and decode instances to launch;
\item the TP degree and baseline frequency of each instance;
\item and the routing weights across instances.
\end{itemize}

To this end, the ILP takes as input a \textit{configuration table} that maps
each candidate prefill or decode instance configuration $c$ to three values:
$
(R_c, E_c, G_c)
$,
where $R_c$ is maximum SLO-feasible goodput, $E_c$ is energy per request, and
$G_c$ is GPU cost. We describe how this table is constructed in
\autoref{subsubsection:configuration_table}.
}
Let $n_c$ denote the number of instances of configuration $c$, $G$ be
total GPUs available, and $\alpha$ be a safety margin to compensate for
modeling and subsampling inaccuracies.
The provisioning problem is formulated as:

\vspace{-12pt}
\begin{align}
\text{minimize} \quad & \sum_c n_c E_c R_c \\
\text{s.t.} \quad 
& \sum_c n_c G_c \le G \\
& \sum_{c \in prefill} n_c R_c \ge (1+\alpha)R \\
& \sum_{c \in decode} n_c R_c \ge (1+\alpha)R \\
& n_c \in \mathbb{N}
\end{align}

The first constraint ensures the total GPU usage does not exceed cluster
capacity.
The second and third constraints ensure that prefill and decode independently
have sufficient capacity to handle the target goodput with margin $\alpha$.
This separation is critical in disaggregated serving, since SLO violations in
either phase cannot be compensated by over-provisioning the other phase.

The ILP solution determines the number and configuration of prefill and decode
instances, along with their baseline operating frequencies. Routing weights are
then derived proportionally based on instance capacity.

\subsubsection{Configuration Table Construction}
\label{subsubsection:configuration_table}

%

A key input to the ILP is the configuration table, which maps each candidate
instance configuration $c$ to a tuple $(G_c, R_c, E_c)$ representing GPU cost,
maximum SLO-feasible goodput, and energy per request
for a given input trace. Constructing this table
is challenging because these quantities depend on workload-dependent batching,
queueing, and frequency-sensitive execution dynamics, which are difficult to
capture using closed-form analytical models.

To address this, \name uses a frequency- and memory-aware, iteration-level
inference simulator. The simulator is built on top of offline-trained latency
and power models (\autoref{subsubsec:models}) and accurately reproduces batched inference execution
under realistic workload traces. The simulator takes as input
(1) a request trace,
(2) the instance type (prefill or decode),
(3) tensor-parallel (TP) degree, and
(4) GPU frequency.
Given these inputs, it simulates per-iteration batching, scheduling, and execution while tracking
request-level TTFT and TPOT, iteration-level power consumption, and KV cache usage.
This allows \name to determine both SLO
feasibility and memory feasibility for each configuration under the given workload.

Using this simulator, \name computes the configuration table entries as follows.

%

\paragraph{GPU cost $G_c$.}
For a configuration $c$ with tensor-parallel degree $TP_c$, each shard occupies
one GPU. Thus, the total GPU cost of one instance is:
$G_c = TP_c$.
This formulation naturally extends to other parallelism schemes (e.g., pipeline
parallelism) by accounting for additional GPU requirements.

\paragraph{Maximum sustainable goodput $R_c$.}
To compute $R_c$ (the maximum SLO-feasible RPS for a given trace),
we perform a binary search over candidate request rates. For each candidate
rate, we generate a scaled version of the input trace and run the simulator to
evaluate SLO compliance. If any request violates its SLO, the rate is reduced;
otherwise, it is increased. The maximum feasible rate is recorded as $R_c$.


To generate scaled traces, we use down-sampling (randomly dropping requests)
rather than time dilation. Down-sampling preserves realistic arrival patterns
and variability after load balancing, whereas time dilation artificially alters
temporal correlations and leads to unrealistic per-instance workload dynamics.

\paragraph{Energy per request $E_c$}
For each simulated run (\eg at the maximum sustainable goodput $R_c$), we
estimate energy using our power model. The simulator produces an
iteration-by-iteration execution timeline. For each simulated iteration (batch)
$t$, we predict the average GPU power $P_t$ based on the batch features, TP
degree, and frequency, and obtain the simulated iteration latency $L_t$ from the
latency model. Let $N_{\mathrm{req}}$ denote the number of requests completed in
the simulation. We compute the average energy per request as:
$
E_c = \frac{\sum_t P_t \cdot L_t}{N_{\text{req}}}.
$
For prefill, we additionally account for idle energy during gaps between
batches using an idle-power estimate, which is important because prefill load
can be bursty and instances may temporarily idle between batches.


\subsubsection{Runtime Request Routing}

At runtime, requests arriving at the serving cluster
are routed proportionally to each instance’s maximum
sustainable goodput calculated for the current configuration.
For prefill, we approximate request load using prompt length and distribute
requests accordingly. For decode, we treat requests as having uniform workload
and route based on goodput capacity.
This routing strategy ensures that the burstiness experienced by each instance
aligns with the simulator’s assumptions, preserving Tier 1’s SLO guarantees.
~

\subsection{Tier 2: Fine-Grained DVFS Control}

Tier~1 provisions sufficient capacity to satisfy SLOs under peak load. However,
real workloads exhibit substantial short-term variability: instantaneous
request arrivals and batch compositions fluctuate even when long-term average
load remains stable. Under such conditions, GPUs may operate at unnecessarily
high frequencies during transient dips, resulting in avoidable energy waste.

Tier~2 addresses this inefficiency by dynamically adjusting GPU frequency at
iteration-level granularity. Its objective is to exploit short-term slack while
preserving the SLO guarantees ensured by Tier~1. Because prefill and decode
exhibit fundamentally different dynamics (\autoref{sec:challenges}), we adopt a
stage-aware control strategy.

\textbf{Stage-Aware DVFS Strategy.}
As mentioned in \autoref{sec:challenges}, prefill and decode differ along two
critical dimensions.
First, prefill is compute-bound and highly sensitive to SM frequency, whereas decode is typically memory-bandwidth-bound and less responsive to frequency scaling.
Second, prefill load is arrival-driven and bursty, while decode load evolves
smoothly because requests persist across many iterations.
These differences motivate distinct control mechanisms for the two stages: we
apply model predictive control (MPC) for prefill and a simplified per-batch
policy for decode.

\subsubsection{Prefill: MPC-Based Frequency Control.}
For prefill instances, queueing delay and execution latency jointly determine
TTFT.
Therefore, frequency decisions affect not only the current batch but also
subsequent queue evolution.
As a result, myopic per-batch tuning is insufficient, and we therefore adopt an
MPC approach over a finite horizon of $N$ future batches.
Specifically, at each batch boundary, the controller performs three steps:

\begin{enumerate}[leftmargin=1em,noitemsep,topsep=0pt]
\item Batch projection.  Using the current waiting and running requests, along
    with each request's input lengths, we simulate how they will be scheduled
    into the next $N$ batches.

\item Frequency evaluation.  For each candidate frequency assignment over the
    horizon, we use the latency model to predict batch completion times and
    derive projected TTFT for all active requests.

\item Feasible energy minimization.  Among assignments that satisfy TTFT SLO
    constraints, we select the one with minimum projected energy consumption.
\end{enumerate}

We approximate future uncertainty by assuming (i) no new arrivals within the
horizon and (ii) ongoing requests do not finish within the horizon.
We empirically find these approximations do not significantly affect the
resulting system behavior.
Additionally, since the controller is invoked at every batch boundary, these
approximations are corrected continuously.

\textbf{Efficient Frequency Search}
The MPC formulation requires selecting a frequency vector over $N$ future
batches. With $K$ candidate frequencies, exhaustive search requires evaluating
$K^N$ assignments, which is infeasible in millisecond-scale control loops.
To make MPC practical, we employ the greedy expansion algorithm shown in
\autoref{algo:greedy}.

\begin{algorithm}
\small
\caption{Greedy Frequency Selection}
\label{algo:greedy}
\KwIn{List of $N$ available frequencies $\mathit{freqs\_avail}$, number of future batches $K$}
\KwOut{Frequency assignments $\mathit{freqs\_opt}$ for each future batch in the horizon}

\tcp{Initialize all windows with the highest freq}
$\mathit{freqs\_opt} \gets$ [max(freqs\_avail)] repeated $K$ times\; \label{algline:init}

\tcp{
Iteratively expand the search space by including next two successive freqs
}
\For{$i = 2$ \KwTo $N-1$}{  \label{algline:expand}
    $\mathit{candidates} \gets \emptyset$\;
    
    \tcp{
        Mutate each usage of $\mathit{freqs\_avail}[i-1]$ to
        $\mathit{freqs\_avail}[i]$ and $\mathit{freqs\_avail}[i+1]$. There are $3^{K'}-1$ such mutations, where
        $K'$ is the number of batches using $\mathit{freqs\_avail}[i-1]$
    }
    \ForEach{valid mutation from $\mathit{freqs\_opt}$}{ \label{algline:mutate}
        $\mathit{freqs\_mutate} \gets mutate(freqs\_opt)$\;
        
        \If{meets\_slo($freqs\_mutate$)}{
            Append $\mathit{freqs\_mutate}$ to $\mathit{candidates}$\;
        }
    }
    
    \tcp{
        If exists candidate meeting SLO, pick the lowest average power one;
        otherwise, exit early
    }
    \If{$\mathit{candidates} \ne \emptyset$}{ \label{algline:choose-optim}
        $\mathit{freqs\_opt} \gets \arg\min\limits_{c \in \mathit{candidates}} \frac{\sum_{i=1}^{K} \mathit{latency(c,i)} \cdot \mathit{power(c,i)}}{\sum_{i=1}^{K} \mathit{latency(c,i)}}$
    }
    \Else{
        \KwBreak
    }
}
\Return{$\mathit{freqs\_opt}$}
\end{algorithm}

The algorithm begins by assigning the maximum frequency to all $N$ batches in the
window (\autoref{algline:init}).
It then incrementally explores lower-frequency options by progressively
expanding the search space, including two additional lower frequencies at each
step (\autoref{algline:expand}). We chose two instead of one, as empirically we found one would lead to the algorithm being stuck in a local minima, and two gave better results with only a small additional cost.
To include the $i$-th and $(i+1)$-th candidate frequencies,
starting with the current optimal frequency assignment,
\ie a vector of frequencies, one for each of the $N$ batches,
the algorithm generates a set of mutated frequency assignments,
by enumerating all possible ways all occurrences of the $(i-1)$-th frequency in the
vector can be replaced with the $i$-th and $(i+1)$-th frequency
(\autoref{algline:mutate}).
If there are $K'$ occurrences of the $(i-1)$-th frequency, this
process generates $3^{K'} - 1$ candidate frequency assignments.
Each candidate is then evaluated using the latency model to check compliance with
TTFT
SLO, and among the valid candidates, the
one with the lowest power consumption is selected as the new optimal assignment
(\autoref{algline:choose-optim}).
The process repeats until all candidate frequencies have been considered, and
the final optimal frequency assignment is returned.
If no feasible mutation exists at a given level, the search terminates early.

In practice, we select $N=7$ frequencies from the full set supported by the
GPU, and a max future horizon of $K=8$. We empirically found that $K=8$ can
cover the request completion of all waiting requests for our chosen workloads.
Compared to a brute-force approach, the algorithm reduces the worst-case
complexity from $K^N$ to $O(K \cdot 3^N)$, and after some parallelization
optimizations and early exits, the average running time of the algorithm is
about 4 ms.

\subsubsection{Decode: Per-Batch Frequency Selection}

Decode exhibits smoother load dynamics and weaker sensitivity to GPU frequency
compared to prefill. Because each request persists across many iterations, the
decode queue evolves gradually, and frequency decisions have limited impact on
future scheduling. As a result, horizon-based MPC is unnecessary. Instead,
\name applies a lightweight per-batch frequency selection policy.

The decode SLO is specified in terms of TPOT. However,
because output length is unknown at runtime, directly predicting TPOT is
difficult. We therefore use time between tokens (TBT) as a conservative proxy.
Ensuring that the latency of each decode iteration satisfies the TBT constraint
guarantees that the resulting TPOT also satisfies the SLO.

For each upcoming batch, the controller evaluates candidate frequencies in
ascending order and selects the minimum frequency $f$ whose predicted iteration
latency satisfies the TBT constraint. If no candidate frequency satisfies the
constraint, the controller falls back to the maximum frequency to preserve SLO
compliance. This policy requires at most $K$ latency model evaluations per batch,
making it highly efficient for online control.

Frequency scaling also affects memory pressure indirectly. Lower frequencies
increase iteration latency, prolonging request residency in the decode stage.
This increases the number of concurrent requests and expands the KV cache
footprint. To prevent out-of-memory (OOM) conditions, \name enforces a
KV-cache utilization threshold. If cache usage exceeds this threshold, the
controller overrides the energy-optimal frequency and temporarily selects the
maximum frequency to accelerate request completion and reclaim memory.

\subsection{Modeling Infrastructure}
\label{subsubsec:models}

Both Tier~1 (coarse-grained provisioning) and Tier~2 (fine-grained DVFS) rely
on accurate predictions of latency and energy consumption.
Tier~1 uses these predictions to estimate configuration goodput and per-request
energy during simulation, while Tier~2 uses them to evaluate candidate
frequency assignments under SLO constraints.
Since closed-form analytical models cannot capture the complex interaction
between batching, tensor parallelism, frequency scaling, and workload
variability, we adopt a data-driven modeling approach, by training
iteration-level latency and power models offline for each (LLM model, GPU type)
pair.
Offline profiling is performed once and reused during runtime optimization,
enabling fast decision-making without incurring measurement overhead in the
serving path.

\subsubsection{Latency Modeling}

We build separate latency models for prefill and decode, as the two stages
exhibit different performance characteristics.
The latency model predicts the execution time of a single batch (or iteration)
as a function of: (1) number of requests in the batch,
(2) sum, mean, and standard deviation of input lengths,
(3) tensor-parallel (TP) degree,
and (4) GPU SM frequency.
We use histogram gradient boosting trees due to their strong nonlinear modeling
capacity and low inference overhead.

For training, labels are collected by instrumenting the inference engine (vLLM)
to measure per-iteration execution time under different batch shapes and
frequencies.
For inference, for decode, per-iteration latency directly corresponds to Time
Between Tokens (TBT).
For prefill, batch latency contributes to TTFT together with queueing delay,
which is captured by the simulator and MPC projection.

\subsubsection{Power Modeling}

Accurate energy estimation requires predicting power consumption at the same
iteration granularity.
However, power measurement presents additional challenges: GPU power APIs
(e.g., NVML) report averaged power values at coarse time intervals, which do
not align exactly with individual decode iterations.
To address this, we collect averaged power measurements over repeated runs of
similar batch configurations and use them as labels for model training.

For decode, we train a regression model using the same feature set as the
latency model.
We impose a monotonic constraint along the frequency dimension, ensuring that
predicted power increases with frequency, which empirically improves robustness
by removing noise-induced inconsistencies in the training data.

For prefill, we observe that power behavior is well approximated by a
structured interpolation. We therefore construct a three-dimensional lookup
table indexed by
(1) item total input token length in the batch,
(2) TP degree,
and (3) frequency.
We then use linear interpolation between profiled points to estimate power for
intermediate configurations.

We also model the GPU's idle power, since prefill workload is bursty, and
instances may remain idle between batches and consume non-negligible power.
\cut{
\jonny{I think readers with a fresh mind will ask why can't we measure idle
power by simply letting the GPU sit idle and measure it. Do we have an answer
for this? Otherwise I feel the method we measure idle power is a bit adhoc
and we might as well don't mention how we measured it.} 
\yunzhao{It is because the idle power is not a constant as I mentioned in the previous verison of the paper. It dependes on the recent iterations' batch shape}
\omar{maybe it helps to explain that unloaded idle vs idle for a short duration between workloads is different. probably has something to do at the hardware architecture level, like turning off components if idle for longer duration but not for shorter durations to remain responsive.}
We estimate the idle power as the intercept of power-RPS curves collected
during profiling.
For example, Figure \ref{fig:idle_power} shows the power-RPS curves
corresponding to requests with input lengths 128 and 1024.
We find that power-RPS curves corresponding to different input lengths usually
have similar intercept, and we therefore take the average intercept as the idle
power.

\begin{figure}[t]
    \centering
    \includesvg[width=\columnwidth]{figures/idle_power_780mhz.svg}
    \vspace{-24pt}
    \caption{
        Power-RPS curve of a TP4 prefill instance at 780MHz with different
        request input length and batch size 1. The intercept is the estimated
        idle power. \jonny{Can we make this figure 1/2 the height, and in
        PDF format? Taking up too much space right now.}
    }
    \label{fig:idle_power}
\end{figure}
}





\subsection{Practical Considerations}
\label{subsec:practical}
We describe several additional experimental assumptions that affect how
evaluations are conducted:

\begin{itemize}[leftmargin=*,noitemsep,topsep=0pt]
\item \textit{Future Workload Prediction.}
We use a simplified approach which uses the past 5 minute load as the predicted
load for the next 5 minutes.
This approach assumes the workload does not change significantly over time.
Other more sophisticated workload prediction algorithms may bring additional
gains, but are orthogonal to our work.

\item \textit{Configuration Transition.}
%
During the configuration transition which happens every 5 minutes, we need to
tear down the old instances and spin up new instances, without interrupting
the processing of the incoming requests.
In this work, for simplicity, we run each 5-minute trace in isolation and 
report the result over the steady state. 
Building a complete system with minimal transition overheads can be modeled
after~\cite{dynamollm:hpca2024}.

\item \textit{Operating Margins.}
GPU frequency changes using NVML typically incur tens of milliseconds delay and
are not synchronized across TP GPUs.
Additionally, inaccuracies in the latency prediction model result in potential
SLO violations.
To account for these, we add a margin of 5\% for both the GPU frequency
adjustment latency discussed and the target goodput discussed in
\autoref{subsec:tier1}.
We empirically found a margin of 5\% to be the sweet spot between energy
savings and SLO violations.

\item \textit{Tuning for Stability}
Fine-grained DVFS must remain robust to modeling inaccuracies and frequency
transition delays.
To prevent cascading SLO violations, we incorporate various safety margins into
latency thresholds.
For example, if the observed latency exceeds prediction, we immediately revert
to maximum frequency.
Additionally, for prefill, the controller is additionally triggered upon new
request arrivals to respond quickly to bursts.
Together, these mechanisms ensure that energy optimization does not compromise
service reliability.
\end{itemize}








\section{Implementation}
\label{sec:impl}
We implemented \name on top of vLLM~\cite{kwon2023efficient} in Python, in 1930 lines of code.
The latency and power models are implemented in Python on top of
Scikit-learn~\cite{scikitlearn:research2011} and then converted to ONNX for faster inference.
The simulator is implemented based on DistServe's simulator, which we added
additional functionality to model batch latencies at different frequencies and
power modeling.
We leverage NVIDIA's NVML library~\cite{nvml} to perform the actual frequency
scaling.
To monitor the power draw, we run a separate process that asynchronously
samples GPU power at fixed 10 ms intervals, using the
\verb|NVML_FI_DEV_POWER_INSTANT| API from NVML.

\section{Evaluation}

\begin{figure*}[t]
    \centering
    \begin{subfigure}{0.24\textwidth}
        \centering
        \includegraphics[width=\linewidth]{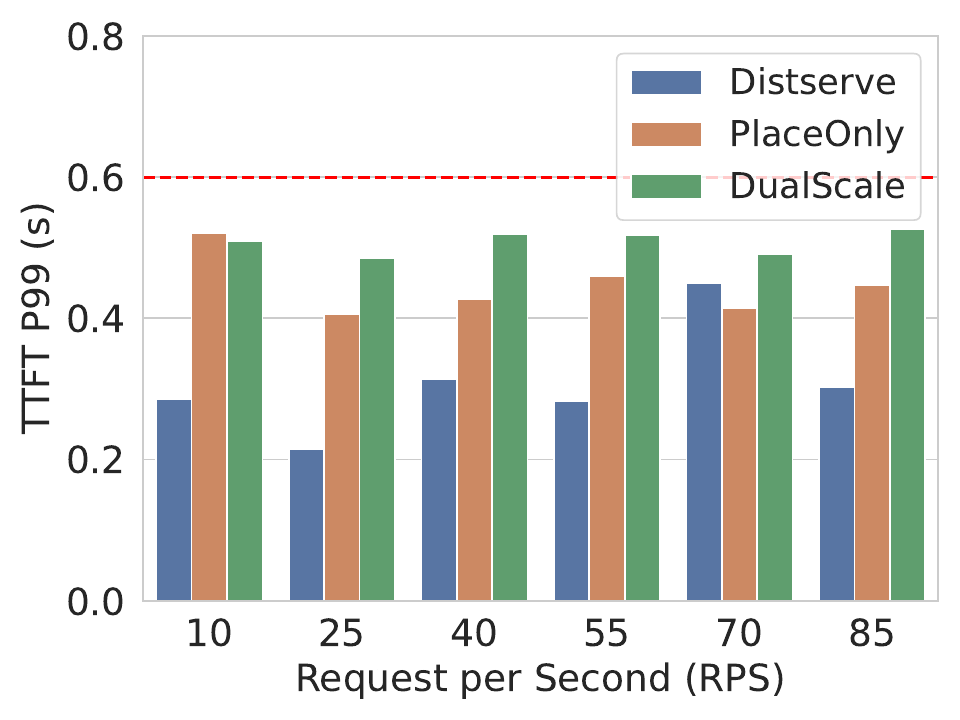}
        \caption{TTFT}
    \end{subfigure}
    \hfill
    \begin{subfigure}{0.24\textwidth}
        \centering
        \includegraphics[width=\linewidth]{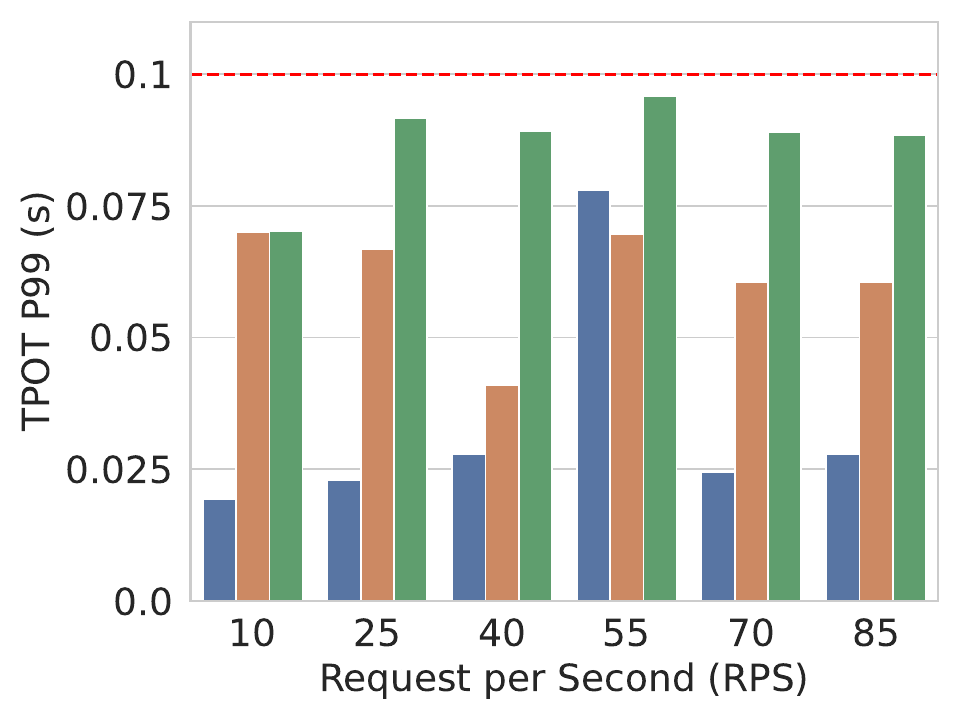}
        \caption{TPOT}
    \end{subfigure}
    \hfill
    \begin{subfigure}{0.24\textwidth}
        \centering
        \includegraphics[width=\linewidth]{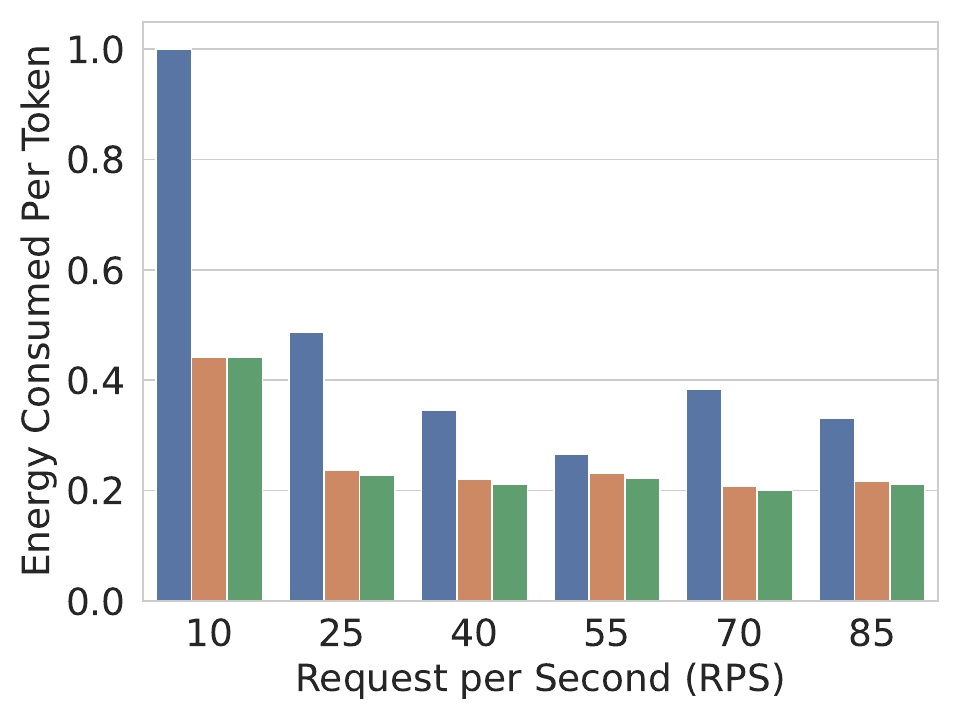}
        \caption{Energy per Token - Decode}
    \end{subfigure}
    \hfill
    \begin{subfigure}{0.24\textwidth}
        \centering
        \includegraphics[width=\linewidth]{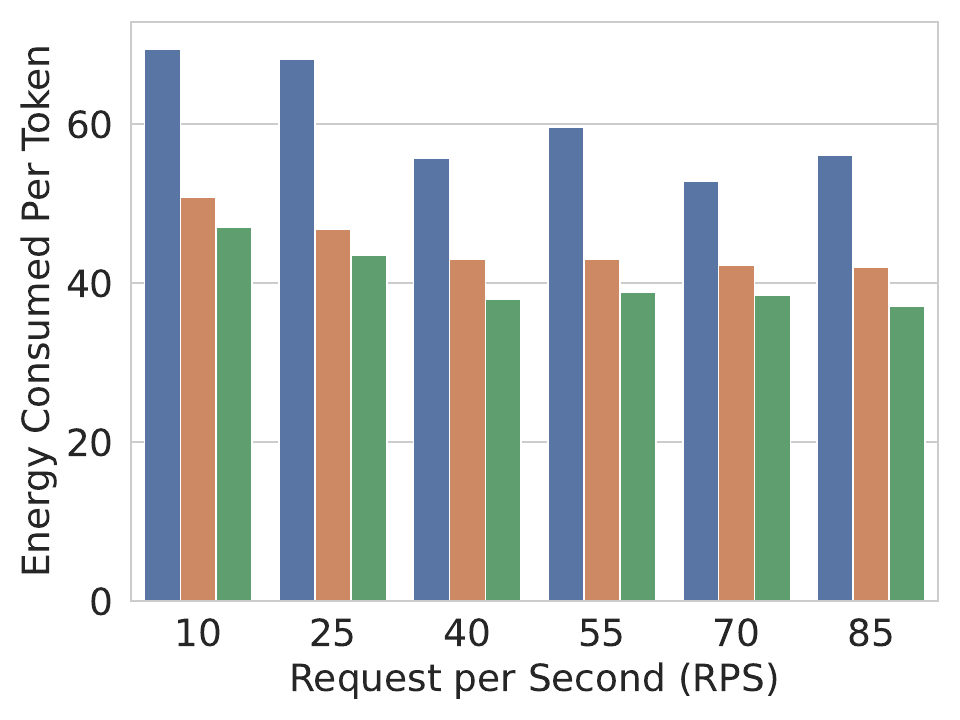}
        \caption{Energy per Token - Prefill}
    \end{subfigure}
    \vspace{-8pt}
    \caption{Results for various controlled workloads with constant average RPS}
    \label{fig:synth-workload}
    \vspace{-8pt}
\end{figure*}

\subsection{Evaluation Methodology}

\parb{Workload}
\label{sec:method-workload}
We construct two complementary workloads: 1) a controlled workload with steady
RPS to evaluate our system under both controlled steady-state conditions, and
2) a realistic time-varying workload to exercise the system's coarse-grained
reconfiguration logic.
For the controlled workload, we generate bursty arrivals by sampling
inter-arrival times from a Gamma distribution with shape parameter 0.5 and a
fixed average RPS.
Having realistic bursty arrivals while maintaining a fixed average RPS enables
controlled comparison across placement and DVFS strategies.
For the realistic workload, following prior model serving
works~\cite{alpaserve:osdi2023,splitwise:isca2024,dynamollm:hpca2024,aladdin:2024arxiv,slo-aware:2024letter,serverlessllm:osdi2024,distserve:osdi,past:asplos2025},
we use request timestamps from a real-world production trace -- the Azure LLM
inference trace~\cite{azure-public-dastaset}, which exhibits multi-timescale
burstiness (\autoref{subsec:workload_dyanmics}), and sequentially assign these
timestamps to ShareGPT requests.
For both workloads, we use ShareGPT~\cite{sharegpt4v:eccv2024} to obtain
representative input and output length distributions.

We scale each trace to a target workload level, \ie average RPS, to study system behavior under
different load intensities.
For the controlled workload, the Gamma distribution mean is directly set to the
desired RPS.
For the Azure-based workload, we apply time dilation to match the target
average rate, which preserves the temporal structure of the original trace.
This allows us to compare system behavior across a range of controlled and
realistic loads.

Target workload levels are derived from measured capacity of the cluster
under a serving system -  we choose \name as we expect it outperform other baselines.
%
First, we derive the cluster's full capacity as the maximum RPS that can be
sustained without violating TTFT or TPOT SLOs, which we obtain by performing
binary search over RPS using a long request trace of 200k requests.
Specifically, at each binary search iteration, we run the coarse-grained
placement algorithm to derive a cluster configuration, and use it to serve the
requests with the fine-grained DVFS system enabled, and check end-to-end SLO
compliance.
Second, after determining the cluster's capacity above,
we evaluate each baseline at 67\% and 85\% of
this value.
Operating below peak capacity is consistent with production practice at
Meta~\cite{meta-tail-utilization} and Microsoft~\cite{patel2024characterizing},
where clusters are typically run with headroom to absorb burstiness and
transient load spikes.

\cut{
Following prior model serving
works~\cite{alpaserve:osdi2023,splitwise:isca2024,dynamollm:hpca2024,aladdin:2024arxiv,slo-aware:2024letter,serverlessllm:osdi2024,distserve:osdi,past:asplos2025},
we use real-world production LLM inference traces, specifically the
ShareGPT~\cite{sharegpt4v:eccv2024} and Azure Public Dataset
~\cite{azure-public-dastaset}.
Since the ShareGPT dataset does not contain request timestamps, we either add
synthetic timestamps to simulate bursty arrival or sequentially assign Azure
request timestamps to each ShareGPT request.
\comment{the flow below is still confuing -- not clear which are about 2 ways of
  sharegpt and which is about hwo to use azure pub trace??}
For the synthetic bursty arrival, we use arrival times sampled from a gamma
distribution of burstiness 0.5~\jonny{is there anything we can cite for 0.5?},
and set the arrival rate of the gamma distribution to our desired RPS.
With the Azure LLM trace, we scale the trace to the desired RPS.

Since running the system at a lower capacity is a common industry practice, we
choose 67\% and 85\% of the full cluster capacity as the target workload.
For example, Meta operates servers at 65-70\% utilization to leave headroom to
absorb bursty traffic~\cite{meta-tail-utilization}, and Microsoft reports that
LLM inference clusters reach peak power utilization of 79\% on average due to
short-term load fluctuations~\cite{patel2024characterizing}.
Specifically, we
determine the full capacity by running experiments with a
200000-request trace \omar{I don't think we used that many, Yunzhao confirm} at different request rates.
The full capacity is the highest request rate the system can handle without SLO
violations.
After determining the full capacity, we scale down the trace by time dilation
to the target workload.
}

\parb{GPU \& LLM Model}
All experiments are conducted on two nodes on Nebius cloud~\cite{nebius}, each
equipped with 8 NVIDIA H100 GPUs connected via InfiniBand, for a total of 16
GPUs.
As the serving model, we use Llama 3.3 70B~\cite{Meta2024}, a widely used
mid-scale Llama model.

\parb{Metrics and SLOs}
Following standard practice in LLM serving, we report TTFT and TPOT as latency
metrics.
For TTFT, we report the 99th percentile (P99) across requests.
For TPOT, we first compute the mean TPOT per request and then report the P99
across requests.
We set the TTFT SLO to 600 ms and the TPOT SLO to 100 ms. The TTFT target is
derived from the SLO used in DistServe for a similarly sized model, adjusted to
reflect performance differences between A100 and H100 GPUs.

\cut{
Specifically, we compute the mean TPOT per request and report the 99th
percentile (P99) across all requests.
For TTFT, we directly report the P99 across requests.
We choose 600 ms as the TTFT SLO and 100 ms as the TPOT SLO.
The TTFT SLO is based on Distserve's TTFT SLO of a similarly sized model,
scaled according to the difference in GPU for A100 vs H100.
}

For power draw, we report the average GPU power (in watts) over the duration of
each session, summed across all GPUs.
For energy efficiency, we report normalized energy in joules per token.
Specifically, prefill energy is normalized by the number of processed requests
(each request produces one first token), and decode energy is normalized by the
total number of generated output tokens.
%

To ensure measurement consistency, we exclude the ramp-up and ramp-down phases
of each session and compute all metrics using only the steady-state portion.
We define the ramp-up region as the period between session start and 30 seconds
after session start, which we empirically find sufficient for the system to
reach steady state. We define the ramp-down region as the period between the
end of request issuance and session end (i.e., when all in-flight requests are
drained). 

\parb{Baselines} We compare \name against the following baselines:

\begin{itemize}[leftmargin=*,noitemsep,topsep=0pt]
    \item \textbf{{\distserve}}: This baseline follows the configuration
        strategy described in~\cite{distserve:osdi}, which generates a
        placement that satisfies TTFT and TPOT SLOs while maximizing throughput
        per GPU. All GPUs operate at the maximum supported frequency.

    \item \textbf{{\namebaseline}}: This baseline applies only the Tier~1
        coarse-grained provisioning described in \autoref{subsec:tier1},
        selecting a placement that minimizes energy consumption subject to SLO
        constraints. Unlike DistServe, it allows GPUs to operate at fixed
        frequencies that may be lower than the maximum. However, it does not
        perform fine-grained DVFS during execution.
\end{itemize}

\subsection{End-to-end Results}
\label{subsec:e2e-results}

We conduct end-to-end evaluations of \name using two workloads described in
\autoref{sec:method-workload}: a controlled workload and a realistic
time-varying workload.

%

\begin{table}[t]
\centering
\caption{Cluster placement for 5--10 minute period. Note that \name uses the
same cluster configuration as \namebaseline, so it is not listed separately.}
\label{tab:prod-config-small}
\vspace{-8pt}

\resizebox{\columnwidth}{!}{
\begin{tabular}{c c c c c c}
\toprule
\textbf{Load} & \textbf{System} & \textbf{Phase} & \textbf{TP} & \makecell{\textbf{Frequency}\\\textbf{(GHz)}} & \makecell{\textbf{Load balancing}\\\textbf{weights (\%)}} \\
\midrule

\multirow{4}{*}{67\%}
 & \multirow{2}{*}{DistServe}
   & Prefill & $3\times TP2$ & 1.83, 1.83, 1.83 & 33.3, 33.3, 33.3 \\
 & & Decode  & $2\times TP4$ & 1.83, 1.83 & 50.0, 50.0 \\

 & \multirow{2}{*}{\namebaseline}
   & Prefill & $4\times TP2$ & 1.08, 1.08, 1.35, 1.23 & 22.2, 22.2, 29.3, 26.3 \\
 & & Decode  & $2\times TP4$ & 1.08, 1.35 & 46.0, 54.0 \\

\midrule

\multirow{4}{*}{85\%}
 & \multirow{2}{*}{DistServe}
   & Prefill & $4\times TP2$ & 1.83, 1.83, 1.83, 1.83 & 25.0, 25.0, 25.0, 25.0 \\
 & & Decode  & $2\times TP4$ & 1.83, 1.83 & 50.0, 50.0 \\

 & \multirow{2}{*}{\namebaseline}
   & Prefill & $4\times TP2$ & 1.83, 1.83, 1.32, 1.32 & 26.8, 26.8, 23.2, 23.2 \\
 & & Decode  & $2\times TP4$ & 1.56, 1.47 & 52.0, 48.0 \\

\bottomrule
\end{tabular}
\vspace{-8pt}
}
\end{table}

\subsubsection{Controlled Workload}
\label{sec:e2e_const}

%
We generate the controlled workload following
\autoref{sec:method-workload}, varying the target RPS from 10 to 85 in
increments of 15 RPS.
The resulting cluster configurations are listed in Appendix
\autoref{tab:prod-configs}, and the TTFT, TPOT, and energy consumption for both
prefill and decode are reported in \autoref{fig:synth-workload}.

\name satisfies TTFT and TPOT SLOs across all evaluated RPS values.
As shown in the two left plots of \autoref{fig:synth-workload}, P99 TTFT and
TPOT remain below their respective SLO thresholds (indicated by red lines).
Compared to the two baselines, \name typically exhibits higher P99 latency values
while still meeting SLOs.
This reflects its DVFS strategy, which elongates iterations as much as possible
without violating latency constraints in order to reduce energy consumption.

In terms of energy efficiency, DistServe consistently consumes the highest
energy per token.
For decode, \namebaseline and \name achieve comparable energy consumption.
For prefill, \name achieves the lowest energy consumption, followed by \namebaseline and then DistServe.
Specifically, \namebaseline reduces prefill energy by 20–31\% relative to
DistServe, while \name achieves a larger reduction of 27–36\%. We analyze the
underlying causes of these differences in
\autoref{subsec:energy_saving_analysis}.

\begin{figure*}[tp]
    \centering
    
    \begin{subfigure}[t]{0.42\textwidth}
        \centering
        \includegraphics[width=\linewidth]{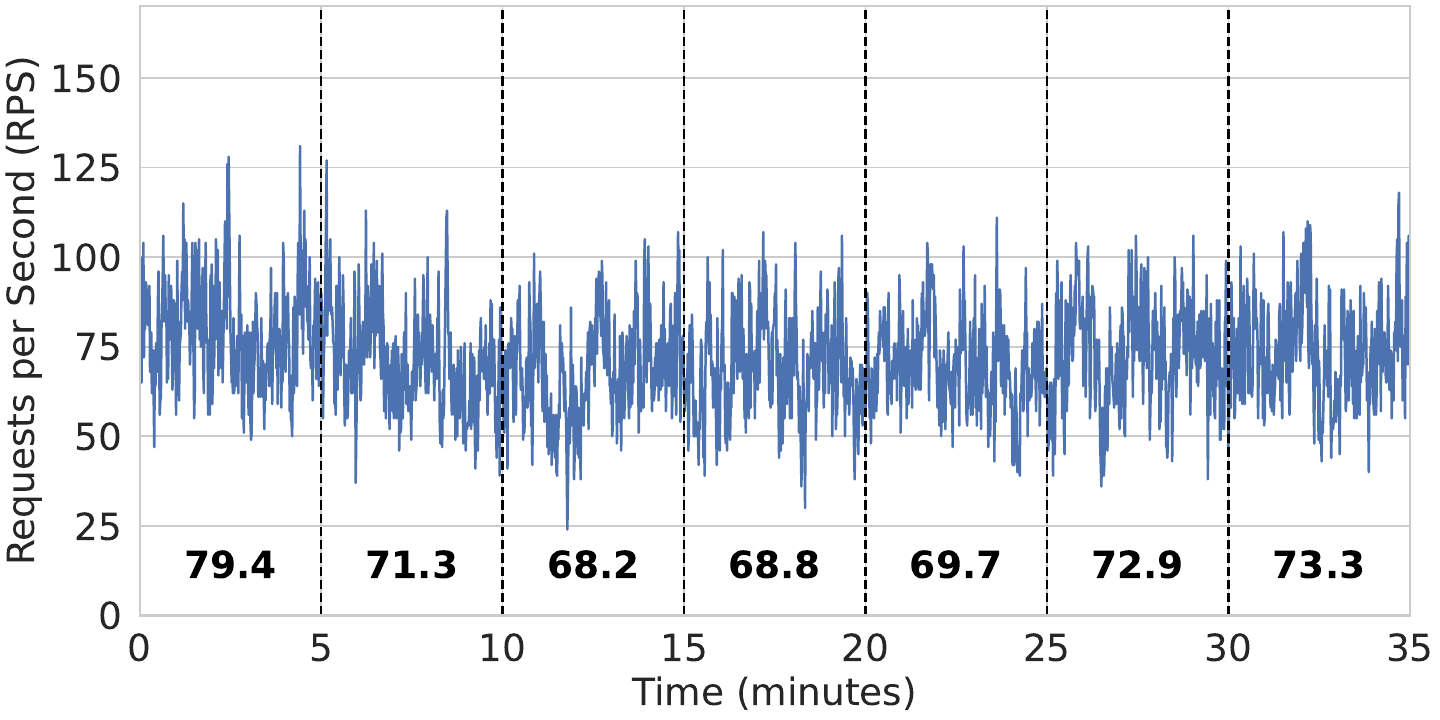}
        \caption{Workload scaled to 67\% of the full capacity.}
    \end{subfigure}
    \hspace{0.04\textwidth}
    \begin{subfigure}[t]{0.42\textwidth}
        \centering
        \includegraphics[width=\linewidth]{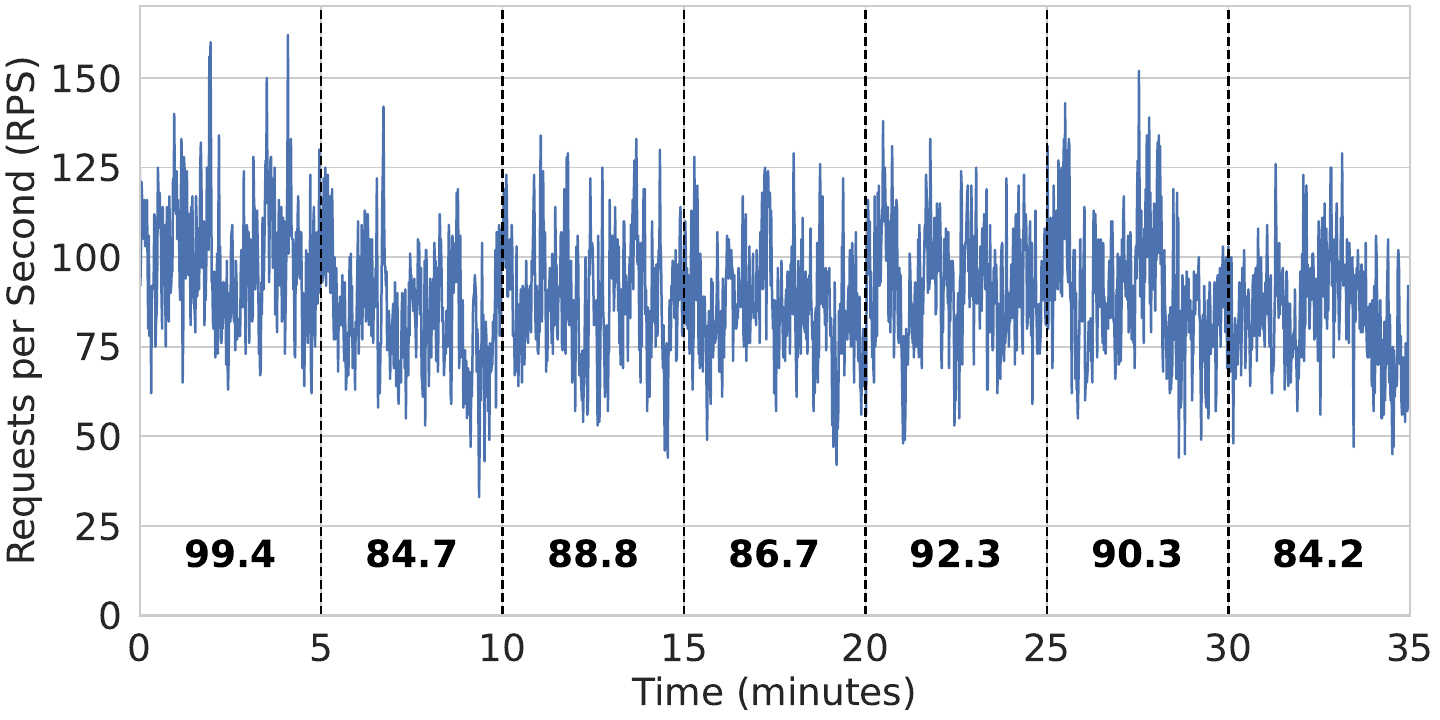}
        \caption{Workload scaled to 85\% of the full capacity.}
    \end{subfigure}
    \vspace{-6pt}
    \caption{The production workload in terms of RPS, divided into 5-minute chunks.}
    \label{fig:rps}
\end{figure*}

\begin{figure*}[t]
    \centering
    \begin{subfigure}{0.24\textwidth}
        \centering
        \includegraphics[width=\linewidth]{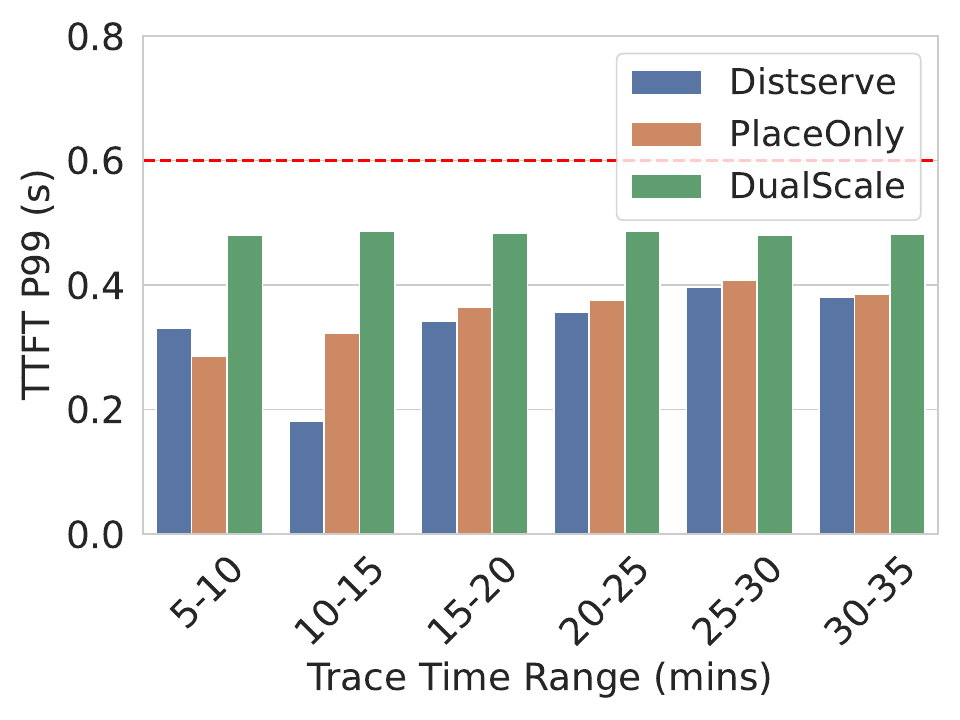}
        \vspace{-16pt}
        \caption{TTFT}
    \end{subfigure}
    \hfill
    \begin{subfigure}{0.24\textwidth}
        \centering
        \includegraphics[width=\linewidth]{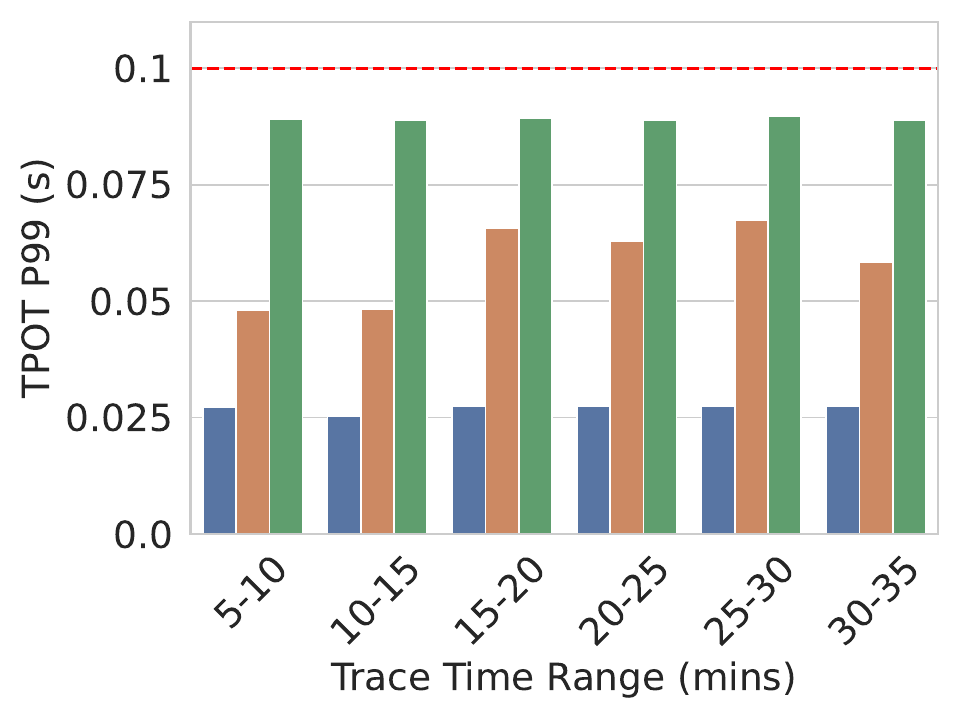}
        \vspace{-16pt}
        \caption{TPOT}
    \end{subfigure}
    \hfill
    \begin{subfigure}{0.24\textwidth}
        \centering
        \includegraphics[width=\linewidth]{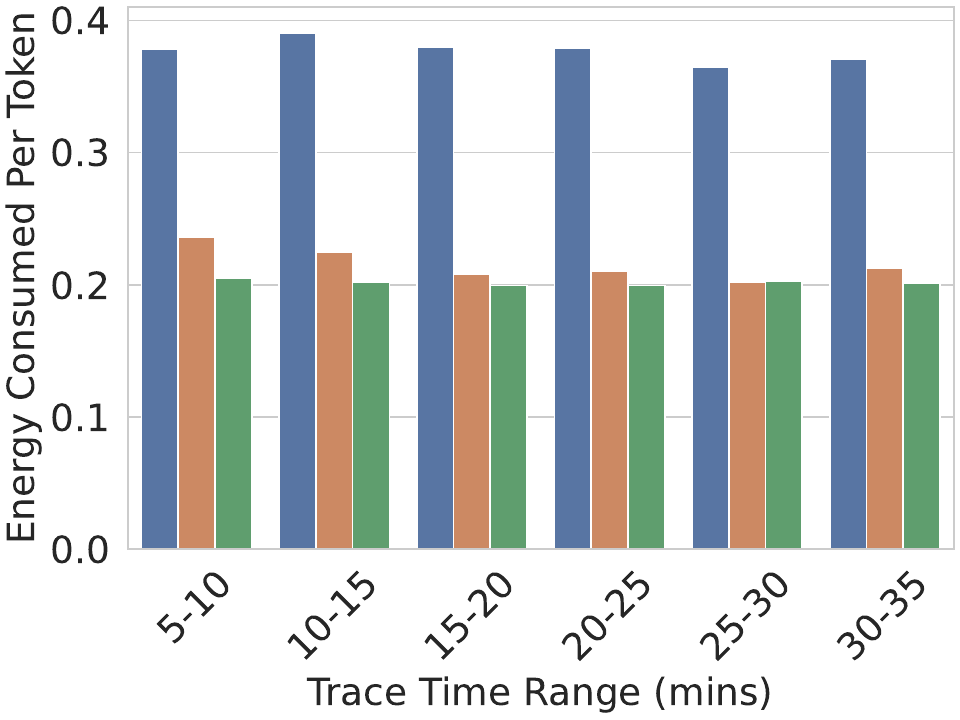}
        \vspace{-16pt}
        \caption{Energy per token - Decode}
    \end{subfigure}
    \hfill
    \begin{subfigure}{0.24\textwidth}
        \centering
        \includegraphics[width=\linewidth]{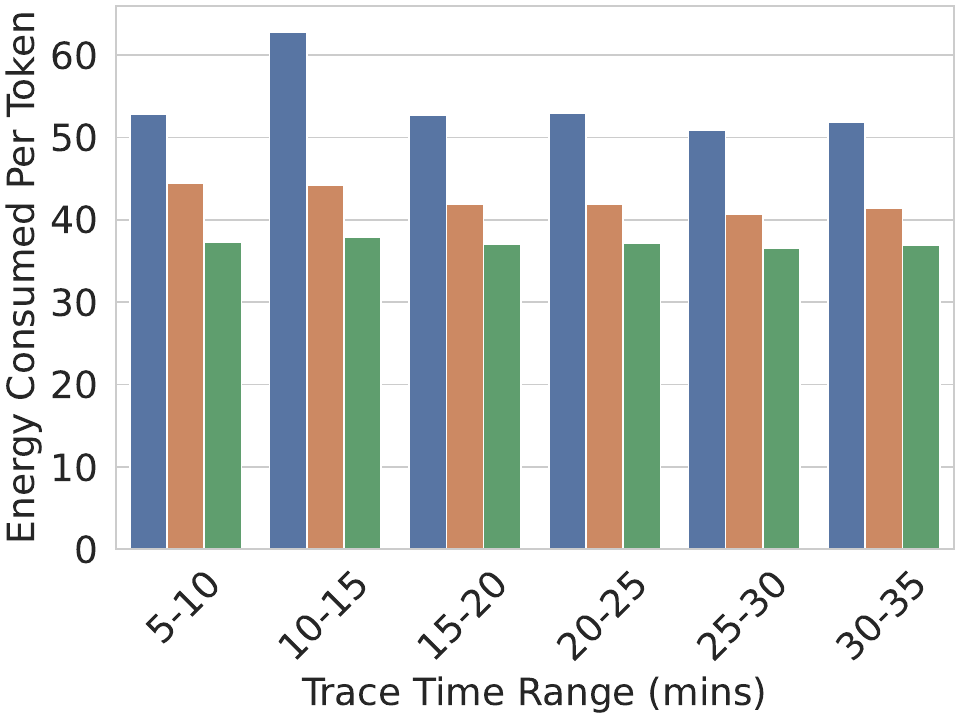}
        \vspace{-16pt}
        \caption{Energy per token - Prefill}
    \end{subfigure}

    \vspace{0.8em}

    \begin{subfigure}{0.24\textwidth}
        \centering
        \includegraphics[width=\linewidth]{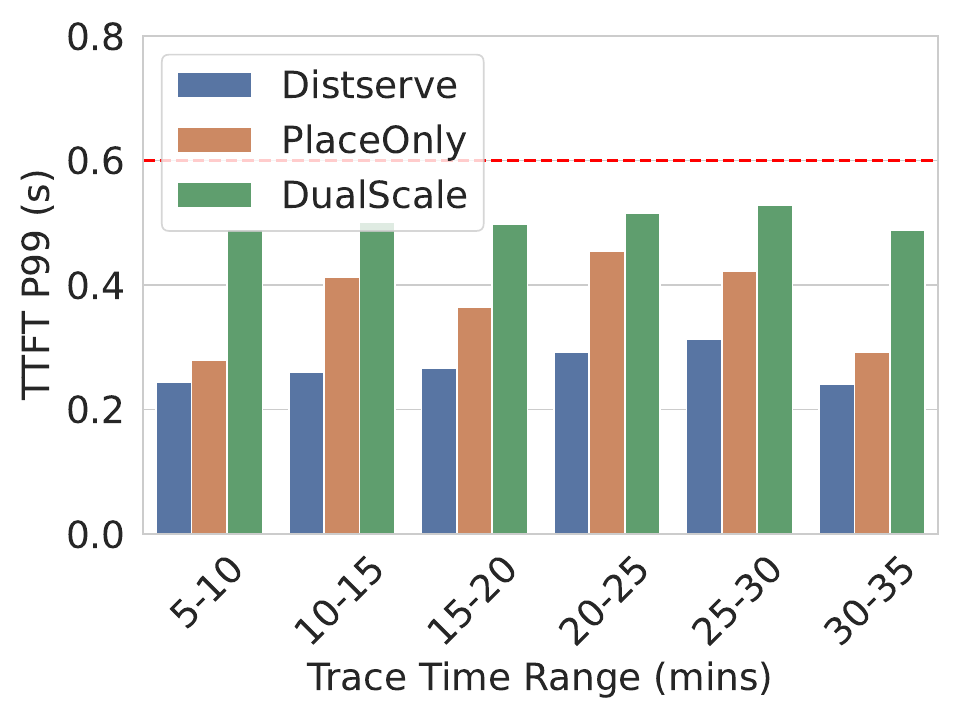}
        \vspace{-16pt}
        \caption{TTFT}
    \end{subfigure}
    \hfill
    \begin{subfigure}{0.24\textwidth}
        \centering
        \includegraphics[width=\linewidth]{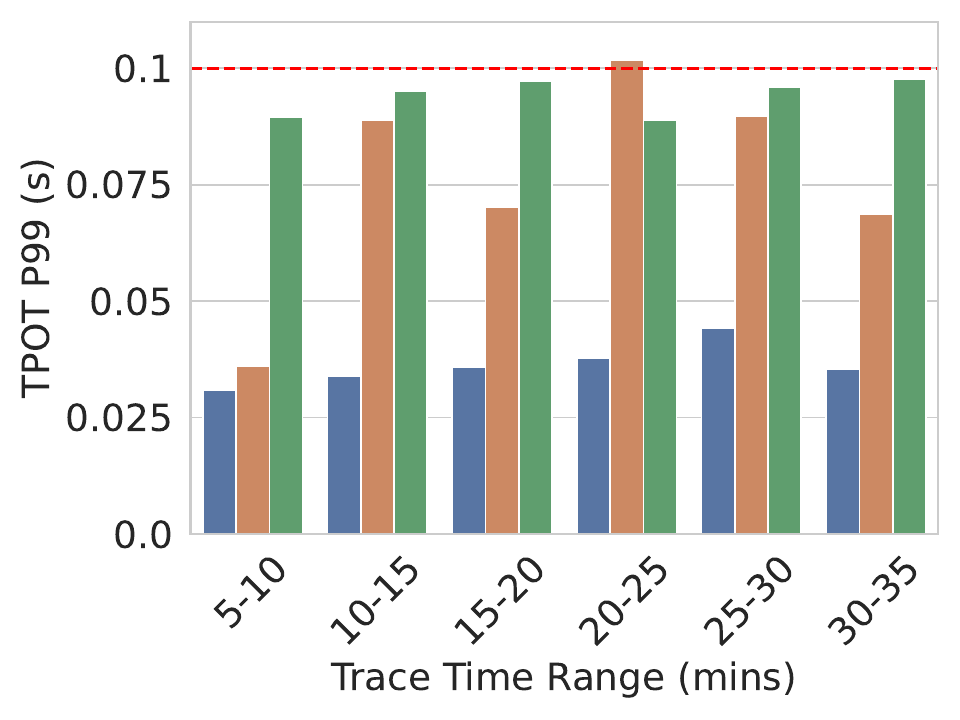}
        \vspace{-16pt}
        \caption{TPOT}
    \end{subfigure}
    \hfill
    \begin{subfigure}{0.24\textwidth}
        \centering
        \includegraphics[width=\linewidth]{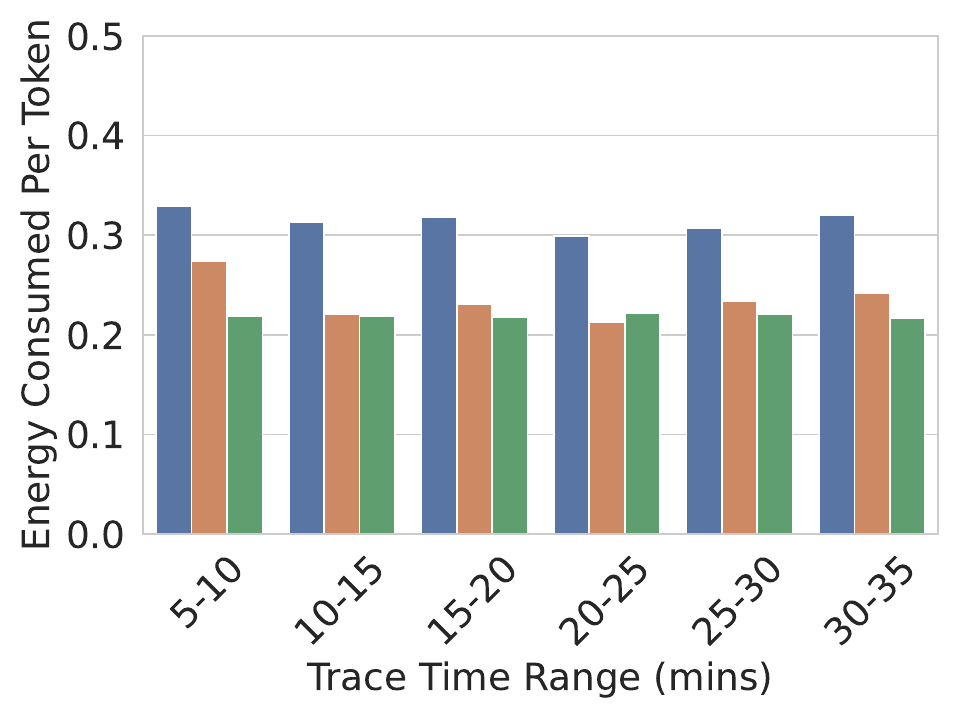}
        \vspace{-16pt}
        \caption{Energy per token - Decode}
    \end{subfigure}
    \hfill
    \begin{subfigure}{0.24\textwidth}
        \centering
        \includegraphics[width=\linewidth]{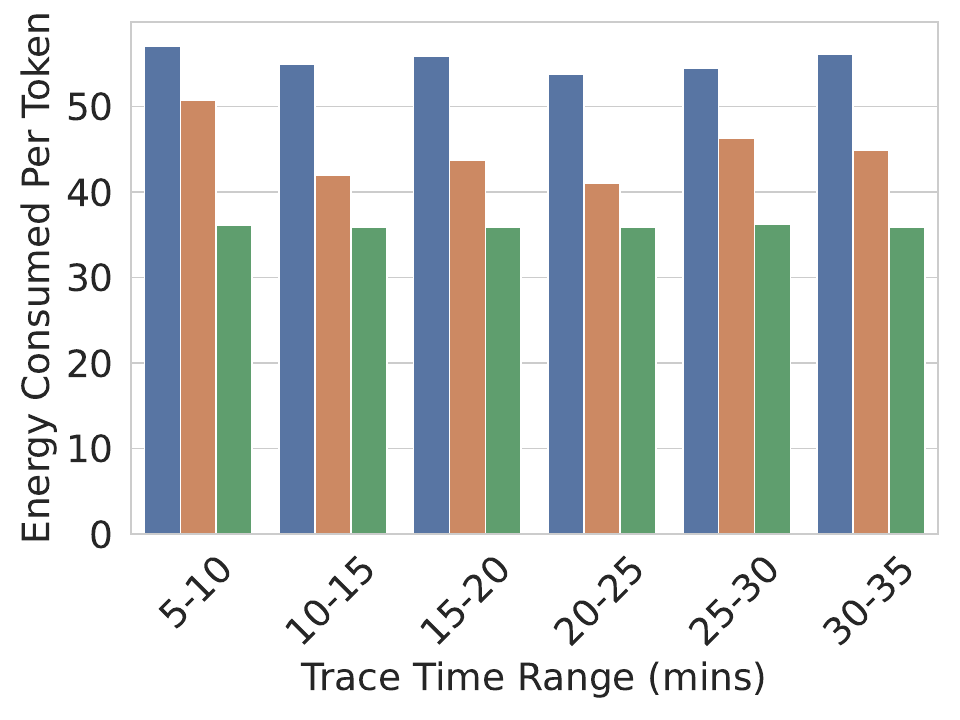}
        \vspace{-16pt}
        \caption{Energy per token - Prefill}
    \end{subfigure}

    \caption{Results for 30 minute long production traces at 67\% (1st row) and 85\% (2nd row) capacity of the system}
    \label{fig:prod-workload}
    \vspace{-8pt}
\end{figure*}

\subsubsection{Production Workload}
\label{subsubsec:production_workload}

We next evaluate the systems using the realistic workload described in
\autoref{sec:method-workload}, scaled to 67\% and 85\% of full capacity as
described in \autoref{sec:method-workload}.
This workload captures realistic, time-varying request arrivals and exercises
the system's coarse-grained reconfiguration mechanism. The scaled traces are
shown in \autoref{fig:rps}, where vertical lines denote 5-minute intervals at
which Tier~1 reconfiguration occurs.
Since each configuration is generated based on the load observed in the
preceding 5-minute window, we exclude the initial 0–5 minute interval and
report results starting from minute 5.
The configuration for the first evaluated window (minutes 5–10) is shown in
\autoref{tab:prod-config-small}, while subsequent configurations are listed in
Appendix \autoref{tab:prod-configs}.

The results are shown in \autoref{fig:prod-workload}, where the top and
bottom rows correspond to 67\% and 85\% workload, respectively.
We observe that all systems satisfy TTFT and TPOT SLOs under both load levels,
with one minor exception: under the 85\% workload, \namebaseline exceeds the TPOT
SLO by 1.5 ms during the 20–25 minute window.
Consistent with the controlled workload setting, \name operates closer to the
SLO boundaries, exhibiting higher P99 TTFT and TPOT values while remaining
within latency constraints.

Energy trends are consistent between the 67\% and 85\% workload settings.
For decode, DistServe consumes the most energy, while \namebaseline and \name
achieve comparable reductions.
Under 67\% load, \namebaseline reduces decode energy by 37–45\% relative to
DistServe, while \name achieves a 44–48\% reduction.
In contrast, for prefill, \name consistently achieves the lowest energy consumption,
followed by \namebaseline and then DistServe.
Under 67\% load, \namebaseline reduces prefill energy by 16–29\% relative to
DistServe, while \name achieves larger reductions of 28–39\%.
Similar trends are observed at 85\% load.

\cut{
For the energy per token consumption of the 85\% workload, we again see the same pattern for both decode and prefill. \namebaseline has 16\% to 29\% lower energy consumption than DistServe, while \name has 26\% to 34\% lower energy consumption than DistServe for decode instances.
In decode, there is one 5-minute period,
\comment{which one?}
where \namebaseline has lower energy consumption than \name, but it must also be noted that the \namebaseline system could not meet the TPOT SLO for this period.
And for prefill, \namebaseline has 10\% to 23\% lower energy consumption than DistServe, and \name has 33\% to 36\% lower energy consumption.

In summary, comparing DistServe to \namebaseline, we show that static
configuration selection alone accounts for significant energy savings.
By comparing \namebaseline to \name, we further show that incorporating
fine-grained DVFS further optimizes prefill energy consumption; however, DVFS's
energy saving is much less in the decode phase, especially for the controlled
workload case. \comment{what is the other vcase? this sounds like case
dependent and hence a bit arbitray} \jonny{I think the production workload
case. Hoping the revised writing earlier makes this clear. Or I think we can
just not compare controlled v.s. production workload here?}
}


\subsection{Energy Saving Analysis}
\label{subsec:energy_saving_analysis}

\begin{figure}[t]
    \centering
    \includegraphics[width=0.8\columnwidth]{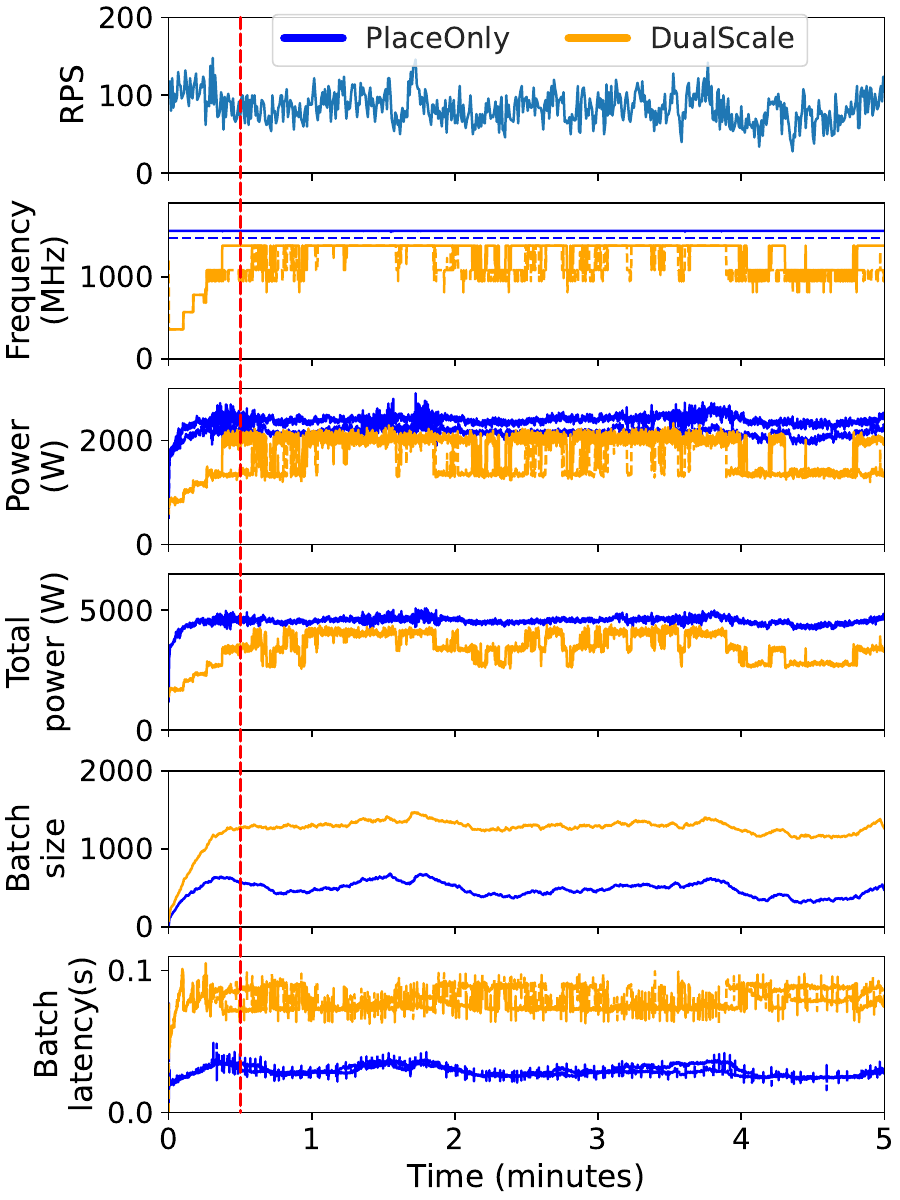}
    \vspace{-8pt}
    \caption{
        Decode over-configuration case (5-10 minutes 85\% capacity workload).
    }
    \label{fig:combo-decode-1}
    \vspace{-8pt}
\end{figure}

\begin{figure}[t]
    \centering
    \includegraphics[width=0.8\columnwidth]{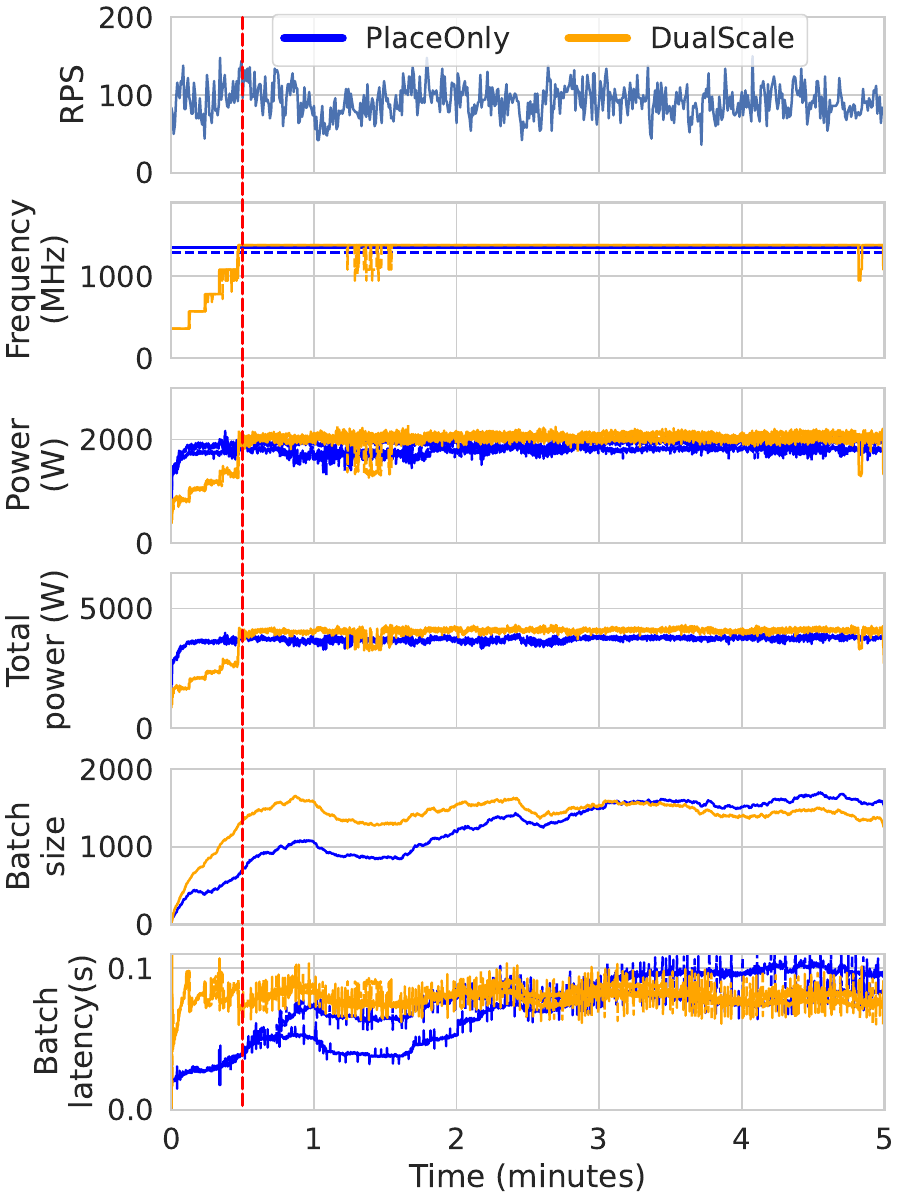}
    \vspace{-8pt}
    \caption{Decode under-configuration case (20-25 minutes 85\% capacity workload).}
    \label{fig:combo-decode-2}
    \vspace{-8pt}
\end{figure}

\begin{figure}[t]
    \centering
    \includegraphics[width=0.8\columnwidth]{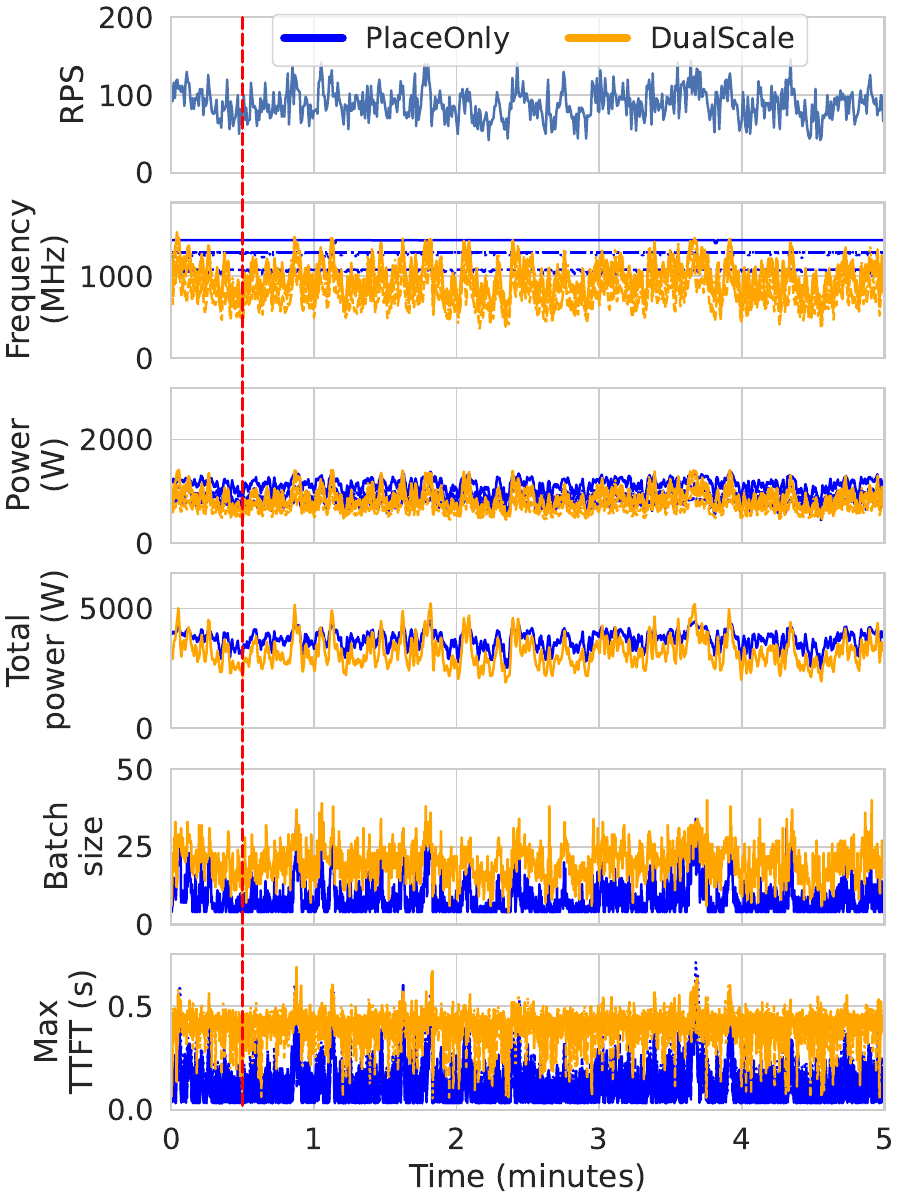}
    \vspace{-8pt}
    \caption{Prefill phase timelines in the 10-15 minute window, under 85\%
    capacity workload.}
    \label{fig:combo-prefill}
    \vspace{-8pt}
\end{figure}

\subsubsection{Impact of Workload Prediction}

We next analyze where the additional savings of \name (over \namebaseline) come
from. \autoref{fig:synth-workload} and \autoref{fig:prod-workload} show that
most savings are obtained by selecting an energy-efficient static configuration
for each workload window (i.e., \namebaseline), which reduces the energy of
DistServe by 11\% to 29\% (20\% on average) for prefill, and 16\% to 45\% (33\%
on average) for decode.
DVFS provides a smaller incremental gain: ranging from -4\% to 20\% (6\% on
average) for decode, and 9\% to 29\% (15\% on average) for prefill.
%
Note that the -4\% energy gain is from a 5-minute window where
\namebaseline violated the SLO.
%
Hence, most of the incremental savings from DVFS come from prefill, which
contributes about ${2.5\times}$ more than decode.

Comparing controlled and production traces clarifies the role of workload
prediction in placement.
Under controlled traces, decode savings over \namebaseline are only 2.8\% (vs.
8.6\% for prefill), whereas under production traces they increase to 6.3\% for
decode (and 15.5\% for prefill).
In controlled traces, \namebaseline is derived from the known trace (for the
current window) and is therefore close to optimal, leaving little headroom for
DVFS.
In production traces, we predict each next 5-minute workload from the previous
5-minute window; this introduces both over-prediction and under-prediction
(\autoref{fig:rps}).
DVFS is therefore useful as a correction mechanism that
compensates for prediction error online, which explains the larger gains under
realistic workloads.

To further localize these effects, we next examine individual 5-minute windows
and compare system dynamics for decode and prefill. This window-level analysis
reveals where the savings come from and why DVFS yields smaller gains for
decode than for prefill.

\subsubsection{Decode}

We first analyze decode at the granularity of individual 5-minute windows to
explain why DVFS has lower energy-saving potential than in prefill. We focus
on two representative windows from the 85\% workload: (i) an over-configured
window (5--10 minutes), where \name achieves clear savings over \namebaseline, and
(ii) an under-configured window (20--25 minutes), where \name consumes more
energy than \namebaseline.

\autoref{fig:combo-decode-1} and \autoref{fig:combo-decode-2} show internal
dynamics for these two windows. Each figure includes six time-series plots:
input RPS, per-instance frequency (two instances), per-instance decode power,
total GPU power, batch size, and batch latency.
%
%
As discussed in \autoref{sec:method-workload}, we treat the first 30 seconds as
ramp-up and exclude it from energy and latency metrics.

In the over-configured window (\autoref{fig:combo-decode-1}), \namebaseline
selects higher fixed frequencies for both decode instances than \name.
%
%
This over-provisioning comes from load over-prediction in the previous window,
so \namebaseline runs at unnecessarily high fixed frequencies. In contrast,
\name applies DVFS online to lower frequency when slack exists, mitigating the
energy penalty of over-provisioning. As a result, \name draws less power over
the window and consumes less total energy.

In the under-configured window (\autoref{fig:combo-decode-2}), both
\namebaseline and \name configuration is determined by the lower RPS in the
previous 5-minute window and is therefore too under-provisioned for the current
demand.
Since \namebaseline operates at the fixed pre-determined frequency, it cannot
raise the frequency to account for under-provisioning, and eventually violates
the TPOT SLO during minutes 4-5.
In contrast, with the same under-provisioned cluster configuration, \name
dynamically adjust the frequency based on transient workload change, and
operates at a slightly higher frequency than \namebaseline, to meet the SLO.

\cut{
A third case, not shown due to space constraints, exhibits nearly identical
energy per token for \namebaseline and \name; relative to the under-configured case
in \autoref{fig:combo-decode-2}, the main difference is that \name has
slightly lower power than \namebaseline.
}

In summary, decode savings depend on whether the static configuration over- or
under-provisions the next 5-minute window. When \namebaseline over-provisions,
\name lowers frequency and energy; when \namebaseline under-provisions, \name may
raise frequency to protect SLOs, reducing gains.

\subsubsection{Prefill}

For prefill, as was shown in \autoref{subsubsec:production_workload}, \name
achieves the lowest energy consumption in all 5-minute windows. We therefore
present one representative window (10--15 minutes) in
\autoref{fig:combo-prefill}, using the same timeline views as in the decode
analysis.

Immediately, we observe that compared to decode, prefill batch size tracks
input workload much more closely; bursty RPS directly translates to bursty
batches.
This variation in compute demand drives frequency and thus power variations in
\name, while \namebaseline keeps a fixed frequency.
During low-load periods, \name can reduce frequency aggressively to save
energy, while during bursts or backlog growth it can temporarily raise
frequency to protect SLOs.
As a result, \name exhibits a more adaptive power profile and lower overall
energy. The maximum TTFT under \name also stays closer to the SLO target,
indicating that \name reduces energy while still meeting latency constraints.

\subsection{Microscopic View of DVFS}

\begin{figure}[t]
    \centering
    \includegraphics[width=\columnwidth]{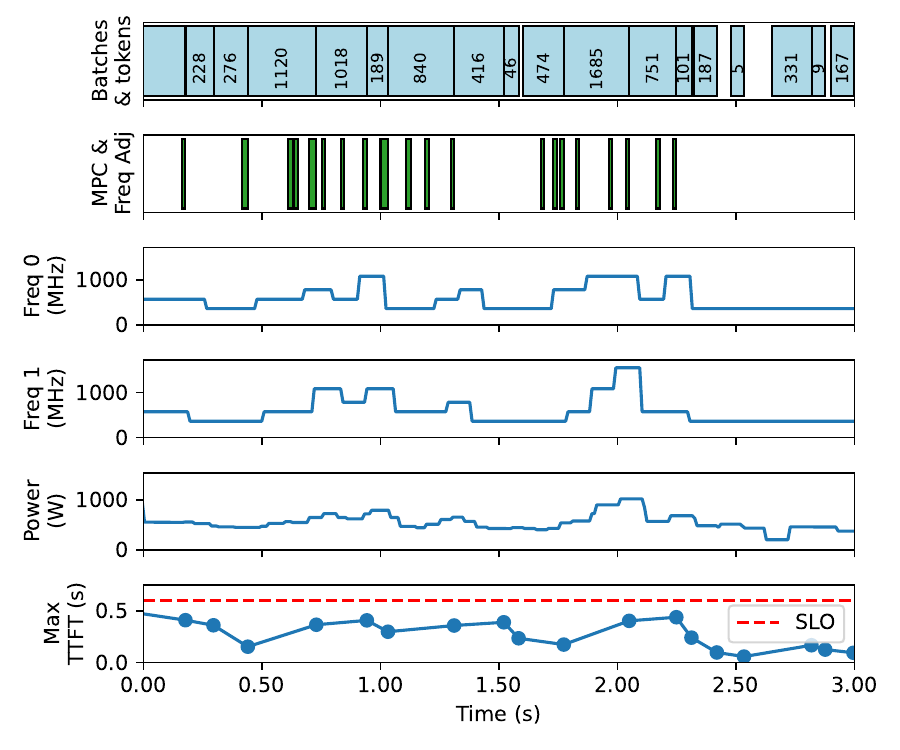}
    \vspace{-24pt}
    \caption{Microscopic behavior of the prefill instance.}
    \label{fig:micro_prefill}
    \vspace{-4pt}
\end{figure}

\begin{figure}[t]
    \centering
    \includegraphics[width=\columnwidth]{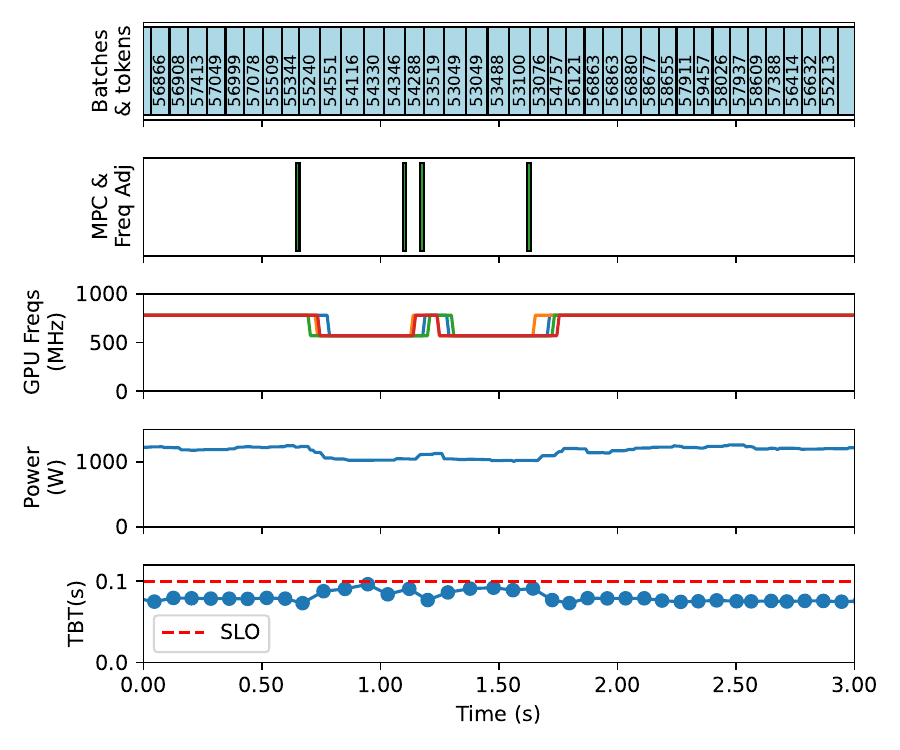}
    \vspace{-24pt}
    \caption{Microscopic behavior of the decode instance. In
    the top figure, each request count as N tokens, where N is the running
    request length.}
    \label{fig:micro_decode}
    \vspace{-4pt}
\end{figure}

After analyzing 5-minute timelines, we next zoom in to sub-second behavior to
show how \name operates online for prefill and decode in
\autoref{fig:micro_prefill} and \autoref{fig:micro_decode}, respectively.
In both figures, from top to bottom, we show:
(1) individual batches as rectangles annotated with batch token count;
(2) MPC events as rectangles, where each rectangle marks one MPC/DVFS
invocation (we only plot invocations that changed frequency to avoid clutter);
(3-4) frequency of each TP rank, where prefill uses a TP 2 instance, and decode
uses TP 4;
(5) total system power draw summed across prefill GPUs or decode GPUs for each
case;
and (6) per-batch latency metric, \ie TTFT for prefill and TBT for decode, with
one dot per batch.

\parb{Prefill}
When the load is low, \eg before 0.5 s during batches of shape 228 and 276, 
\name applies the minimum frequency and power drops
accordingly. As request pressure increases, queueing grows and \name raises
frequency, 
\eg from 360 MHz to 570 and then 780 MHz during the batch of shape 474,
all of which were the lowest frequency that could satisfy the SLO; the
maximum TTFT remains below the 600 ms target.
Most frequency updates occur at batch boundaries or when enough new requests
arrive during an executing batch.
%
We also observe that frequency sometimes change within batches
(\autoref{subsec:practical}) due to new request arrivals, \eg as seen in the
batch of shape 474 at around 1.7 seconds.
%
Additionally, we observe an under-prediction event around 2 seconds (the batch
of size 1685): the executing batch runs longer than predicted, \name thus
immediately switches to maximum frequency to mitigate the effect of
under-prediction, and then lowers frequency again at the next batch boundary.

One important detail is that GPUs in an instance do not always follow
identical frequency trajectories, and observed frequency changes can lag MPC
invocation times.
This is expected for two reasons: frequency application latency (discussed in
\autoref{subsec:practical}) and boost-frequency behavior (frequencies above
1095 MHz are boost states and are sustained only when thermal and power
headroom permit).
Thermal and power characteristics are dynamic and can vary across GPUs even
within one node, leading to small per-rank differences.

\parb{Decode}
\autoref{fig:micro_decode} shows that frequency changes are much less frequent
than in prefill because decode batch size evolves more gradually over time. As
requests arrive and finish, \name alternates between higher and lower
frequencies to meet the SLO while minimizing energy.
The power trace follows these adjustments closely: lowering frequency reduces
power, and raising frequency restores it. In decode, opportunities to run at
lower frequency are short and intermittent, so the incremental DVFS energy
benefit is correspondingly modest.

\subsection{Latency/Power Model Accuracy}
\begin{figure*}[t]
    \centering
    \begin{subfigure}[t]{0.22\textwidth}
        \centering
        \includegraphics[height=\linewidth]{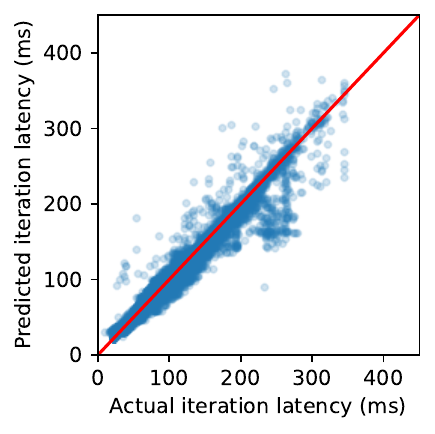}
        \vspace{-20pt}
        \caption{Prefill iteration latency (MAPE=2.9\%).}
        \label{fig:8.4_latency_prefill}
    \end{subfigure}
    \hfill
    \begin{subfigure}[t]{0.22\textwidth}
        \centering
        \includegraphics[height=\linewidth]{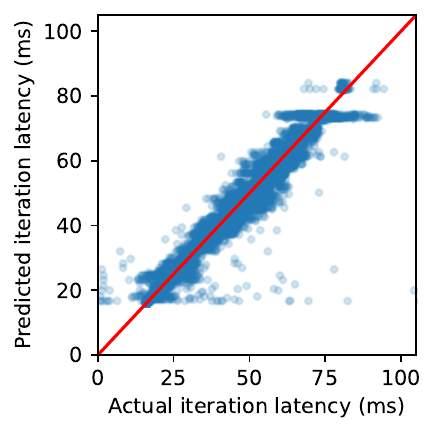}
        \vspace{-20pt}
        \caption{Decode iteration latency (MAPE=2.7\%).}
        \label{fig:8.4_latency_decode}
    \end{subfigure}
    \hfill
    \begin{subfigure}[t]{0.22\textwidth}
        \centering
        \includegraphics[height=\linewidth]{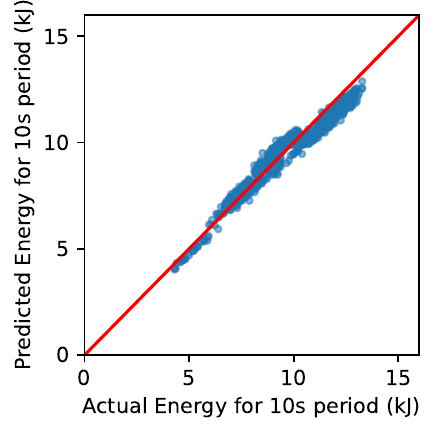}
        \vspace{-20pt}
        \caption{Prefill energy consumption (MAPE=4.1\%).}
        \label{fig:8.4_power_prefill}
    \end{subfigure}
    \hfill
    \begin{subfigure}[t]{0.22\textwidth}
        \centering
        \includegraphics[height=\linewidth]{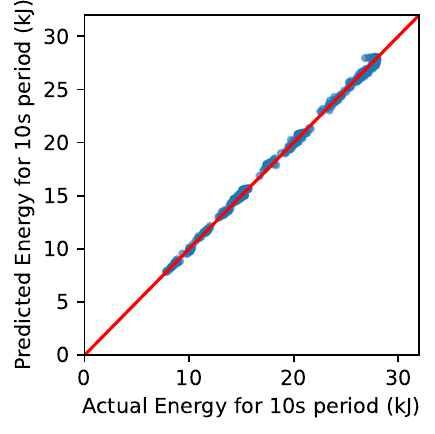}
        \vspace{-20pt}
        \caption{Decode energy consumption (MAPE=1.0\%).}
        \label{fig:8.4_power_decode}
    \end{subfigure}
    \vspace{-8pt}
    \caption{Accuracy of latency and power models against measured values.}
\end{figure*}

\begin{figure*}[t]
    \centering
    \begin{subfigure}[t]{0.22\textwidth}
        \centering
        \includegraphics[height=\linewidth]{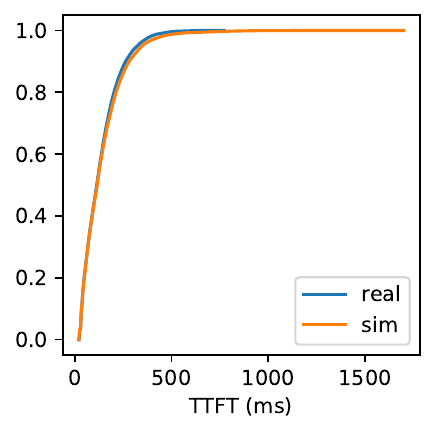}
        \vspace{-20pt}
        \caption{CDF of TTFT.}
        \label{fig:cdf_ttft_sim}
    \end{subfigure}
    \hfill
    \begin{subfigure}[t]{0.22\textwidth}
        \centering
        \includegraphics[height=\linewidth]{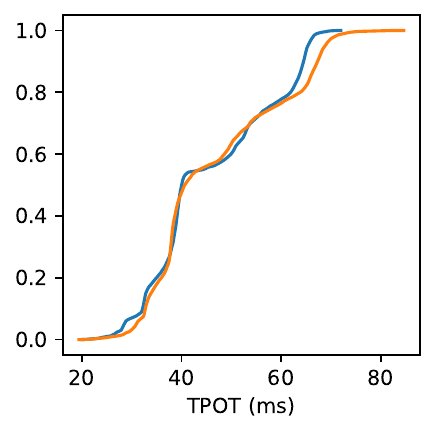}
        \vspace{-20pt}
        \caption{CDF of TPOT.}
        \label{fig:cdf_tpot_sim}
    \end{subfigure}
    \hfill
    \begin{subfigure}[t]{0.22\textwidth}
        \centering
        \includegraphics[height=\linewidth]{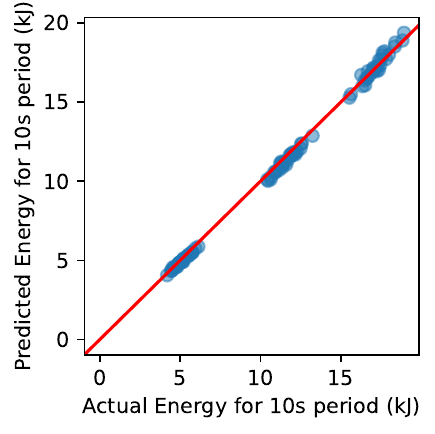}
        \vspace{-20pt}
        \caption{Prefill energy consumption (MAPE=2.3\%).}
        \label{fig:prefill_energy_sim}
    \end{subfigure}
    \hfill
    \begin{subfigure}[t]{0.22\textwidth}
        \centering
        \includegraphics[height=\linewidth]{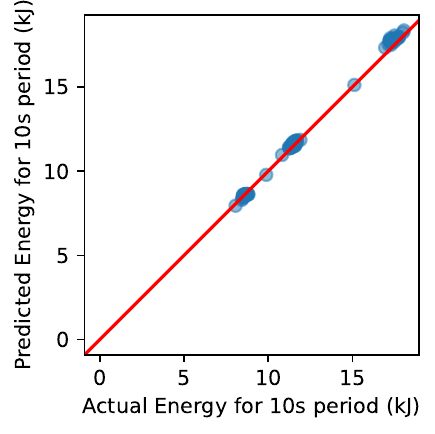}
        \vspace{-20pt}
        \caption{Decode energy consumption (MAPE=1.2\%).}
        \label{fig:decode_energy_sim}
    \end{subfigure}
    \vspace{-8pt}
    \caption{Accuracy of the Tier-1 simulator against real runs, shown by TTFT
    CDF, TPOT CDF, and energy comparisons.}
\end{figure*}

Accurate latency and power models are essential because DVFS decisions and
placement quality both depend directly on these predictions.
We use iteration-level data collected in the constant-frequency experiments
(DistServe and \namebaseline in \autoref{sec:e2e_const}) to evaluate model accuracy.
\autoref{fig:8.4_latency_prefill} and \autoref{fig:8.4_latency_decode} compare
predicted and measured iteration latency.
The latency model achieves mean absolute percentage error (MAPE) of 2.9\% for
prefill instances and 2.7\% for decode instances.

To evaluate the power model, we compare predicted and measured energy over consecutive
10-second windows,
as shown in \autoref{fig:8.4_power_prefill} and \autoref{fig:8.4_power_decode}
for prefill and decode, respectively.
We use this granularity because power is sampled every 100
ms and measurements are often not aligned to iteration boundaries.
Overall, the power model achieves MAPE of 4.1\% for prefill instances and 1.0\% for
decode instances.
These errors are small enough for both online DVFS control and
placement generation.

\subsection{Simulation Accuracy}
Accurate simulation by the simulator is essential because 
the Tier-1 placement algorithm uses it to evaluate candidate
placements before deployment.
%
%
We evaluate simulation accuracy using the 10 RPS, 25 RPS, and 40 RPS traces
from \autoref{sec:e2e_const}. For each trace, the simulator uses the same
frequency set as the placement produced by the Tier-1 placement algorithm.
We then compare simulator outputs against latency and power measurements
collected from the real system under the same traces.
\autoref{fig:cdf_ttft_sim} and \autoref{fig:cdf_tpot_sim} compare the CDFs of
simulated and measured TTFT and TPOT. TTFT CDFs are closely aligned.
We see that TPOT CDFs
are also broadly aligned, with the largest deviation in the top 10\% tail 
where simulation is slightly higher than measurement, which
makes placements mildly conservative while still preserving SLO validity.

We next compare simulated and measured cumulative energy for each consecutive
10-second window.
\autoref{fig:prefill_energy_sim} and \autoref{fig:decode_energy_sim} compare
prefill and decode energy between simulation and the real engine. The simulated
prefill energy achieves a MAPE of 2.3\%, and the simulated decode energy achieves a MAPE of
1.2\%. These results indicate that simulator fidelity is sufficient for Tier-1
placement decisions.

\section{Related Work}

\textbf{Energy-efficient LLM serving.}
Several recent works have explored reducing the energy usage of LLM inference
through placement and parallelism optimizations or GPU frequency tuning.
DynamoLLM~\cite{dynamollm:hpca2024} divides GPUs into pools that serve requests
with similar input and output lengths and optimizes parallelism configurations
within each pool to reduce energy consumption.
However, when computing placements, it assumes that all GPUs within a pool
operate at the same frequency and parallelism settings, rather than modeling
these as decision variables.
throttLL'eM~\cite{throttllem:hpca25} applies predictive GPU throttling to
reduce energy while meeting latency SLOs, but assumes fixed serving placement
and parallelism settings.
Neither system considers prefill--decode disaggregation, where prefill and
decode exhibit distinct burstiness patterns, bottlenecks, and SLO
sensitivities.

\textbf{Energy-efficient LLM training.}
Prior work on energy efficiency in LLM training, such as
Zeus~\cite{you2023zeus} and Perseus~\cite{chung2024reducing}, leverages DVFS to
reduce energy by exploiting iteration-level slack in long-running and
predictable workloads.
These training-oriented systems do not directly transfer to online serving,
which must handle bursty workloads while satisfying strict latency SLOs.

\textbf{General LLM serving systems.}
A large body of work improves LLM serving performance through batching,
scheduling, parallelism, and prefill–decode
disaggregation~\cite{orca:osdi2022,kwon2023efficient,li2023accelerating,agrawal2024taming,song2024powerinfer,lin2024parrot,distserve:osdi,splitwise:isca2024,wu2024loongserve,liu2024cachegen}.
These systems primarily optimize throughput or latency. Our approach is
orthogonal and complementary: it targets energy reduction through DVFS while
preserving serving SLOs.

\section{Conclusion}

In this paper, we presented \name, a two-tier framework for energy-efficient disaggregated LLM serving that jointly optimizes placement and DVFS across prefill and decode while preserving strict TTFT and TPOT SLOs. \name combines model-driven, phase-aware placement at coarse timescales with fine-grained, stage-specific DVFS control to exploit workload slack and correct prediction errors safely.

Evaluation on a multi-GPU cluster using production-style workloads
shows that \name substantially reduces energy consumption compared to
state-of-the-art disaggregated serving systems while consistently
meeting latency SLOs. These results demonstrate that coordinated,
phase-aware control across placement and DVFS is essential for
improving the energy efficiency of disaggregated LLM serving.

\bibliographystyle{ACM-Reference-Format}
\bibliography{dsnl}

@misc{k,
	note = {\url{http://en.wikipedia.org/wiki/K-means\_clustering}},
	title = {k-means clustering}
}

@misc{2,
	note = {http://www.emarketer.com/Article/2-Billion-Consumers-Worldwide-Smartphones-by-2016/1011694},
	title = {2 Billion Consumers Worldwide to Get Smart(phones) by 2016 (Over half of mobile phone users globally will have smartphones in 2018). December 2014}
}

@inproceedings{inferline:socc2020,
	author = {Crankshaw, Daniel and Sela, Gur-Eyal and Mo, Xiangxi and Zumar, Corey and Stoica, Ion and Gonzalez, Joseph and Tumanov, Alexey},
	booktitle = {Proc. of ACM SoCC},
	title = {InferLine: latency-aware provisioning and scaling for prediction serving pipelines},
	year = {2020},
	pages = {477--491}
}

@inproceedings{alpaserve:osdi2023,
	author = {Zhuohan Li and Lianmin Zheng and Yinmin Zhong and Vincent Liu and Ying Sheng and Xin Jin and Yanping Huang and Zhifeng Chen and Hao Zhang and Joseph E. Gonzalez and Ion Stoica},
	booktitle = {Proc. of USENIX OSDI},
	title = {{AlpaServe}: Statistical Multiplexing with Model Parallelism for Deep Learning Serving},
	year = {2023},
	address = {Boston, MA},
	month = {jul},
	pages = {663--679},
	publisher = {USENIX Association},
	isbn = {978-1-939133-34-2},
	url = {https://www.usenix.org/conference/osdi23/presentation/li-zhouhan}
}

@inproceedings{orca:osdi2022,
	author = {Yu, Gyeong-In and Jeong, Joo Seong and Kim, Geon-Woo and Kim, Soojeong and Chun, Byung-Gon},
	booktitle = {Proc. of USENIX OSDI},
	title = {Orca: A distributed serving system for $\{$Transformer-Based$\}$ generative models},
	year = {2022},
	pages = {521--538}
}

@proceedings{TP,
	title = {TP-Link Talon AD7200 Multi-Band Wi-Fi Router},
	url = {http://www.tp-link.com/us/products/details/cat-5506_AD7200.html}
}

@article{scikitlearn:research2011,
	author = {Pedregosa, F. and Varoquaux, G. and Gramfort, A. and Michel, V. and Thirion, B. and Grisel, O. and Blondel, M. and Prettenhofer, P. and Weiss, R. and Dubourg, V. and Vanderplas, J. and Passos, A. and Cournapeau, D. and Brucher, M. and Perrot, M. and Duchesnay, E.},
	journal = {Journal of Machine Learning Research},
	title = {Scikit-learn: Machine Learning in {P}ython},
	year = {2011},
	pages = {2825--2830},
	volume = {12}
}

@misc{Meta2024,
	author = {Meta},
	date = {2024},
	title = {{Meta Llama 3.}},
	howpublished = {\url{https://llama.meta.com/llama3}}
}

@inproceedings{distserve:osdi,
	author = {Zhong, Yinmin and Liu, Shengyu and Chen, Junda and Hu, Jianbo and Zhu, Yibo and Liu, Xuanzhe and Jin, Xin and Zhang, Hao},
	booktitle = {Proc. of USENIX OSDI},
	date = {2024},
	title = {$\{$DistServe$\}$: Disaggregating Prefill and Decoding for Goodput-optimized Large Language Model Serving},
	pages = {193--210}
}

@inproceedings{splitwise:isca2024,
  title={Splitwise: Efficient generative llm inference using phase splitting},
  author={Patel, Pratyush and Choukse, Esha and Zhang, Chaojie and Shah, Aashaka and Goiri, {\'I}{\~n}igo and Maleki, Saeed and Bianchini, Ricardo},
  booktitle={2024 ACM/IEEE 51st Annual International Symposium on Computer Architecture (ISCA)},
  pages={118--132},
  year={2024},
  organization={IEEE}
}

@inproceedings{sharegpt4v:eccv2024,
  title={Sharegpt4v: Improving large multi-modal models with better captions},
  author={Chen, Lin and Li, Jinsong and Dong, Xiaoyi and Zhang, Pan and He, Conghui and Wang, Jiaqi and Zhao, Feng and Lin, Dahua},
  booktitle={European Conference on Computer Vision},
  pages={370--387},
  year={2024},
  organization={Springer}
}

@inproceedings{dynamollm:hpca2024,
  title={Dynamollm: Designing llm inference clusters for performance and energy efficiency},
  author={Stojkovic, Jovan and Zhang, Chaojie and Goiri, {\'I}{\~n}igo and Torrellas, Josep and Choukse, Esha},
  booktitle={2025 IEEE International Symposium on High Performance Computer Architecture (HPCA)},
  pages={1348--1362},
  year={2025},
  organization={IEEE}
}

@misc{azure-public-dastaset,
    author = {Azure},
	note = {\url{https://github.com/Azure/AzurePublicDataset}},
	title = {{Azure Public Dataset}},
	date = {2025},
}

@article{aladdin:2024arxiv,
  title={Aladdin: Joint Placement and Scaling for SLO-Aware LLM Serving},
  author={Nie, Chengyi and Fonseca, Rodrigo and Liu, Zhenhua},
  journal={arXiv preprint arXiv:2405.06856},
  year={2024}
}

@inproceedings{serverlessllm:osdi2024,
  title={$\{$ServerlessLLM$\}$:$\{$Low-Latency$\}$ serverless inference for large language models},
  author={Fu, Yao and Xue, Leyang and Huang, Yeqi and Brabete, Andrei-Octavian and Ustiugov, Dmitrii and Patel, Yuvraj and Mai, Luo},
  booktitle={18th USENIX Symposium on Operating Systems Design and Implementation (OSDI 24)},
  pages={135--153},
  year={2024}
}

@inproceedings{past:asplos2025,
  title={Past-Future Scheduler for LLM Serving under SLA Guarantees},
  author={Gong, Ruihao and Bai, Shihao and Wu, Siyu and Fan, Yunqian and Wang, Zaijun and Li, Xiuhong and Yang, Hailong and Liu, Xianglong},
  booktitle={Proceedings of the 30th ACM International Conference on Architectural Support for Programming Languages and Operating Systems, Volume 2},
  pages={798--813},
  year={2025}
}

@inproceedings{kwon2023efficient,
  title={Efficient memory management for large language model serving with pagedattention},
  author={Kwon, Woosuk and Li, Zhuohan and Zhuang, Siyuan and Sheng, Ying and Zheng, Lianmin and Yu, Cody Hao and Gonzalez, Joseph and Zhang, Hao and Stoica, Ion},
  booktitle={Proceedings of the 29th Symposium on Operating Systems Principles},
  pages={611--626},
  year={2023}
}

@misc{nvml,
	note = {\url{https://developer.nvidia.com/management-library-nvml}},
	title = {{NVIDIA Management Library (NVML)}},
	date = {2025},
}

@inproceedings{inferline:osdi2020,
  title={InferLine: ML Inference Pipeline Provisioning and Management for Tight Latency SLOs},
  author={Crankshaw, Daniel and Wang, Xin and Zhou, Guanyu and Franklin, Michael J and Gonzalez, Joseph E and Stoica, Ion},
  booktitle={14th USENIX Symposium on Operating Systems Design and Implementation},
  pages={283--300},
  year={2020}
}

@article{garrett1994analysis,
  title={Analysis, modeling and generation of self-similar VBR video traffic},
  author={Garrett, Mark W and Willinger, Walter},
  journal={ACM SIGCOMM computer communication review},
  volume={24},
  number={4},
  pages={269--280},
  year={1994},
  publisher={ACM New York, NY, USA}
}

@article{leland2002self,
  title={On the self-similar nature of Ethernet traffic (extended version)},
  author={Leland, Will E and Taqqu, Murad S and Willinger, Walter and Wilson, Daniel V},
  journal={IEEE/ACM Transactions on networking},
  volume={2},
  number={1},
  pages={1--15},
  year={2002},
  publisher={IEEE}
}

@inproceedings{deepspeed:sc2022,
  title={DeepSpeed-MoE: Advancing Mixture-of-Experts Inference and Training to Power Next-Generation AI Scale},
  author={Rajbhandari, Samyam and Rasley, Jeff and Ruwase, Olatunji and He, Yuxiong},
  booktitle={Proceedings of the International Conference for High Performance Computing, Networking, Storage and Analysis (SC)},
  year={2022}
}

@inproceedings{you2023zeus,
  title={Zeus: Understanding and optimizing $\{$GPU$\}$ energy consumption of $\{$DNN$\}$ training},
  author={You, Jie and Chung, Jae-Won and Chowdhury, Mosharaf},
  booktitle={20th USENIX Symposium on Networked Systems Design and Implementation (NSDI 23)},
  pages={119--139},
  year={2023}
}

@inproceedings{chung2024reducing,
  title={Reducing energy bloat in large model training},
  author={Chung, Jae-Won and Gu, Yile and Jang, Insu and Meng, Luoxi and Bansal, Nikhil and Chowdhury, Mosharaf},
  booktitle={Proceedings of the ACM SIGOPS 30th Symposium on Operating Systems Principles},
  pages={144--159},
  year={2024}
}

@inproceedings{agrawal2024taming,
  title={Taming $\{$Throughput-Latency$\}$ tradeoff in $\{$LLM$\}$ inference with $\{$Sarathi-Serve$\}$},
  author={Agrawal, Amey and Kedia, Nitin and Panwar, Ashish and Mohan, Jayashree and Kwatra, Nipun and Gulavani, Bhargav and Tumanov, Alexey and Ramjee, Ramachandran},
  booktitle={18th USENIX Symposium on Operating Systems Design and Implementation (OSDI 24)},
  pages={117--134},
  year={2024}
}

@misc{meta-tail-utilization,
	note = {\url{https://engineering.fb.com/2024/07/10/production-engineering/tail-utilization-ads-inference-meta/?utm_source=chatgpt.com}},
	title = {{Taming the tail utilization of ads inference at Meta scale}},
	date = {2025},
}

@inproceedings{patel2024characterizing,
  title={Characterizing power management opportunities for llms in the cloud},
  author={Patel, Pratyush and Choukse, Esha and Zhang, Chaojie and Goiri, {\'I}{\~n}igo and Warrier, Brijesh and Mahalingam, Nithish and Bianchini, Ricardo},
  booktitle={Proceedings of the 29th ACM International Conference on Architectural Support for Programming Languages and Operating Systems, Volume 3},
  pages={207--222},
  year={2024}
}

@article{energy_benchmark:hotcarbon2025,
author = {Niu, Chenxu and Zhang, Wei and Zhao, Yongjian and Chen, Yong},
title = {Energy Efficient or Exhaustive? Benchmarking Power Consumption of LLM Inference Engines},
year = {2025},
issue_date = {July 2025},
publisher = {Association for Computing Machinery},
address = {New York, NY, USA},
volume = {5},
number = {2},
url = {https://doi.org/10.1145/3757892.3757900},
doi = {10.1145/3757892.3757900},
month = aug,
pages = {56–62},
numpages = {7},
}

@article{lorido2014review,
  title={A review of auto-scaling techniques for elastic applications in cloud environments},
  author={Lorido-Botran, Tania and Miguel-Alonso, Jose and Lozano, Jose A},
  journal={Journal of grid computing},
  volume={12},
  number={4},
  pages={559--592},
  year={2014},
  publisher={Springer}
}

@INPROCEEDINGS{throttllem:hpca25,
  author={Kakolyris, Andreas Kosmas and Masouros, Dimosthenis and Vavaroutsos, Petros and Xydis, Sotirios and Soudris, Dimitrios},
  booktitle={2025 IEEE International Symposium on High Performance Computer Architecture (HPCA)}, 
  title={throttLL’eM: Predictive GPU Throttling for Energy Efficient LLM Inference Serving}, 
  year={2025},
  volume={},
  number={},
  pages={1363-1378},
  doi={10.1109/HPCA61900.2025.00103}}

@ARTICLE{slo-aware:2024letter,
  author={Kakolyris, Andreas Kosmas and Masouros, Dimosthenis and Xydis, Sotirios and Soudris, Dimitrios},
  journal={IEEE Computer Architecture Letters}, 
  title={SLO-Aware GPU DVFS for Energy-Efficient LLM Inference Serving}, 
  year={2024},
  volume={23},
  number={2},
  pages={150-153},
  doi={10.1109/LCA.2024.3406038}}

@inproceedings{song2024powerinfer,
  title={Powerinfer: Fast large language model serving with a consumer-grade gpu},
  author={Song, Yixin and Mi, Zeyu and Xie, Haotong and Chen, Haibo},
  booktitle={Proceedings of the ACM SIGOPS 30th Symposium on Operating Systems Principles},
  pages={590--606},
  year={2024}
}

@inproceedings{li2023accelerating,
  title={Accelerating distributed $\{$MoE$\}$ training and inference with lina},
  author={Li, Jiamin and Jiang, Yimin and Zhu, Yibo and Wang, Cong and Xu, Hong},
  booktitle={2023 USENIX Annual Technical Conference (USENIX ATC 23)},
  pages={945--959},
  year={2023}
}

@inproceedings{lin2024parrot,
  title={Parrot: Efficient serving of $\{$LLM-based$\}$ applications with semantic variable},
  author={Lin, Chaofan and Han, Zhenhua and Zhang, Chengruidong and Yang, Yuqing and Yang, Fan and Chen, Chen and Qiu, Lili},
  booktitle={18th USENIX Symposium on Operating Systems Design and Implementation (OSDI 24)},
  pages={929--945},
  year={2024}
}

@inproceedings{wu2024loongserve,
  title={Loongserve: Efficiently serving long-context large language models with elastic sequence parallelism},
  author={Wu, Bingyang and Liu, Shengyu and Zhong, Yinmin and Sun, Peng and Liu, Xuanzhe and Jin, Xin},
  booktitle={Proceedings of the ACM SIGOPS 30th Symposium on Operating Systems Principles},
  pages={640--654},
  year={2024}
}

@inproceedings{liu2024cachegen,
  title={Cachegen: Kv cache compression and streaming for fast large language model serving},
  author={Liu, Yuhan and Li, Hanchen and Cheng, Yihua and Ray, Siddhant and Huang, Yuyang and Zhang, Qizheng and Du, Kuntai and Yao, Jiayi and Lu, Shan and Ananthanarayanan, Ganesh and others},
  booktitle={Proceedings of the ACM SIGCOMM 2024 Conference},
  pages={38--56},
  year={2024}
}

@article{stojkovic2024towards,
  title={Towards greener llms: Bringing energy-efficiency to the forefront of llm inference},
  author={Stojkovic, Jovan and Choukse, Esha and Zhang, Chaojie and Goiri, Inigo and Torrellas, Josep},
  journal={arXiv preprint arXiv:2403.20306},
  year={2024}
}

@misc{nebius,
	note = {\url{https://nebius.com/}},
	title = {Nebius AI Cloud Platform}
}

@article{liu2024deepseek,
  title={Deepseek-v3 technical report},
  author={Liu, Aixin and Feng, Bei and Xue, Bing and Wang, Bingxuan and Wu, Bochao and Lu, Chengda and Zhao, Chenggang and Deng, Chengqi and Zhang, Chenyu and Ruan, Chong and others},
  journal={arXiv preprint arXiv:2412.19437},
  year={2024}
}

@misc{nvidia-dynamo-disagg-2025,
  author       = {{NVIDIA}},
  title        = {{NVIDIA Dynamo, A Low-Latency Distributed Inference Framework for Scaling Reasoning AI Models}},
  year         = {2025},
  howpublished = {\url{https://developer.nvidia.com/blog/introducing-nvidia-dynamo-a-low-latency-distributed-inference-framework-for-scaling-reasoning-ai-models/}}
}

@misc{vllm-disagg-2025,
  author       = {{vLLM Project}},
  title        = {{Disaggregated Prefill V1}},
  year         = {2025},
  howpublished = {\url{https://docs.vllm.ai/en/latest/features/disagg_prefill.html}}
}

@misc{microsoft-inference-energy-2025,
  author       = {{Microsoft Research}},
  title        = {{The growing energy footprint of AI inference}},
  year         = {2025},
  howpublished = {\url{https://www.microsoft.com/en-us/research/publication/energy-use-of-ai-inference-efficiency-pathways-and-test-time-compute/}}
}

@misc{openai-gpt5-2025,
  author       = {{OpenAI}},
  title        = {{GPT-5}},
  year         = {2025},
  howpublished = {\url{https://openai.com/gpt-5}}
}

@misc{anthropic-models-overview-2025,
  author       = {{Anthropic}},
  title        = {{Claude Models Overview}},
  year         = {2025},
  howpublished = {\url{https://docs.anthropic.com/en/docs/about-claude/models/overview}}
}

@article{kimi-k2-2025,
  title        = {Kimi K2 Technical Report},
  author       = {{Kimi Team}},
  journal      = {arXiv preprint arXiv:2507.20534},
  year         = {2025}
}

@misc{nvidiacoldstart,
	title = {Reducing Cold Start Latency for LLM Inference with NVIDIA Run:ai Model Streamer},
	year = {2026},
	url = {https://developer.nvidia.com/blog/reducing-cold-start-latency-for-llm-inference-with-nvidia-runai-model-streamer/},
}

\begin{table*}[t]
\centering
\caption{TP degree, frequency, and load balancing weights for prefill and
decode for each 5 minute period. Note that \name uses the same cluster
configuration as Min Energy, so it is not listed separately.}
\label{tab:prod-configs}

\resizebox{1.0\textwidth}{!}{
\begin{tabular}{c c c c c c c c}
\toprule
\textbf{Load} &  & 5--10 mins & 10--15 mins & 15--20 mins & 20--25 mins & 25--30 mins & 30--35 mins \\
\midrule

\multirow[c]{10}{*}{67\%}
  & \makecell{Distserve \\ Prefill}
  & \makecell{3$\times$TP2 \\ 1.83,1.83,1.83 \\ 33.3,33.3,33.3}
  & \makecell{4$\times$TP2 \\ 1.83,1.83,1.83,1.83 \\ 25.0,25.0,25.0,25.0}
  & \makecell{3$\times$TP2 \\ 1.83,1.83,1.83 \\ 33.3,33.3,33.3}
  & \makecell{3$\times$TP2 \\ 1.83,1.83,1.83 \\ 33.3,33.3,33.3}
  & \makecell{3$\times$TP2 \\ 1.83,1.83,1.83 \\ 33.3,33.3,33.3}
  & \makecell{3$\times$TP2 \\ 1.83,1.83,1.83 \\ 33.3,33.3,33.3} \\

  & \makecell{Min Energy \\ Prefill}
  & \makecell{4$\times$TP2 \\ 1.08,1.08,1.23,1.35 \\ 22.1,22.1,29.7,26.1}
  & \makecell{4$\times$TP2 \\ 1.08,1.08,1.08,1.29 \\ 23.7,23.7,23.7,28.8}
  & \makecell{4$\times$TP2 \\ 1.08,1.08,1.11,1.11 \\ 24.1,24.1,25.9,25.9}
  & \makecell{4$\times$TP2 \\ 1.08,1.08,1.08,1.08 \\ 25.0,25.0,25.0,25.0}
  & \makecell{4$\times$TP2 \\ 1.08,1.08,1.08,1.11 \\ 24.8,24.8,24.8,25.6}
  & \makecell{4$\times$TP2 \\ 1.17,1.11,1.11,1.05 \\ 27.0,24.8,24.8,23.5} \\
\cmidrule(lr){2-8}
  & \makecell{Distserve \\ Decode}
  & \makecell{2$\times$TP4 \\ 1.83,1.83 \\ 25.0,25.0}
  & \makecell{2$\times$TP4 \\ 1.83,1.83 \\ 25.0,25.0}
  & \makecell{2$\times$TP4 \\ 1.83,1.83 \\ 25.0,25.0}
  & \makecell{2$\times$TP4 \\ 1.83,1.83 \\ 25.0,25.0}
  & \makecell{2$\times$TP4 \\ 1.83,1.83 \\ 25.0,25.0}
  & \makecell{2$\times$TP4 \\ 1.83,1.83 \\ 25.0,25.0} \\

  & \makecell{Min Energy \\ Decode}
  & \makecell{2$\times$TP4 \\ 1.08,1.35 \\ 41.5,58.5}
  & \makecell{2$\times$TP4 \\ 1.17,1.05 \\ 52.1,47.9}
  & \makecell{2$\times$TP4 \\ 1.05,1.05 \\ 50.0,50.0}
  & \makecell{2$\times$TP4 \\ 1.11,1.05 \\ 51.8,48.2}
  & \makecell{2$\times$TP4 \\ 1.08,1.08 \\ 50.0,50.0}
  & \makecell{2$\times$TP4 \\ 1.08,1.17 \\ 48.4,51.6} \\

\midrule
\multirow[c]{10}{*}{85\%}
  & \makecell{Distserve \\ Prefill}
  & \makecell{4$\times$TP2 \\ 1.83,1.83,1.83,1.83 \\ 25.0,25.0,25.0,25.0}
  & \makecell{4$\times$TP2 \\ 1.83,1.83,1.83,1.83 \\ 25.0,25.0,25.0,25.0}
  & \makecell{4$\times$TP2 \\ 1.83,1.83,1.83,1.83 \\ 25.0,25.0,25.0,25.0}
  & \makecell{4$\times$TP2 \\ 1.83,1.83,1.83,1.83 \\ 25.0,25.0,25.0,25.0}
  & \makecell{4$\times$TP2 \\ 1.83,1.83,1.83,1.83 \\ 25.0,25.0,25.0,25.0}
  & \makecell{4$\times$TP2 \\ 1.83,1.83,1.83,1.83 \\ 25.0,25.0,25.0,25.0} \\

  & \makecell{Min Energy \\ Prefill}
  & \makecell{4$\times$TP2 \\ 1.83,1.83,1.32,1.32 \\ 26.8,26.8,23.2,23.2}
  & \makecell{4$\times$TP2 \\ 1.08,1.29,1.29,1.44 \\ 20.2,25.7,25.7,28.5}
  & \makecell{4$\times$TP2 \\ 1.11,1.38,1.38,1.38 \\ 20.7,26.4,26.4,26.4}
  & \makecell{4$\times$TP2 \\ 1.08,1.26,1.38,1.38 \\ 20.7,24.8,27.3,27.3}
  & \makecell{4$\times$TP2 \\ 1.08,1.44,1.5,1.5 \\   20.2,26.5,26.6,26.6}
  & \makecell{4$\times$TP2 \\ 1.44,1.32,1.32,1.32 \\ 26.3,24.6,24.6,24.6} \\
\cmidrule(lr){2-8}
  & \makecell{Distserve \\ Decode}
  & \makecell{2$\times$TP4 \\ 1.83,1.83 \\ 25.0,25.0}
  & \makecell{2$\times$TP4 \\ 1.83,1.83 \\ 25.0,25.0}
  & \makecell{2$\times$TP4 \\ 1.83,1.83 \\ 25.0,25.0}
  & \makecell{2$\times$TP4 \\ 1.83,1.83 \\ 25.0,25.0}
  & \makecell{2$\times$TP4 \\ 1.83,1.83 \\ 25.0,25.0}
  & \makecell{2$\times$TP4 \\ 1.83,1.83 \\ 25.0,25.0} \\

  & \makecell{Min Energy \\ Decode}
  & \makecell{2$\times$TP4 \\ 1.56,1.47 \\ 52.0,48.0}
  & \makecell{2$\times$TP4 \\ 1.11,1.47 \\ 43.6,56.4}
  & \makecell{2$\times$TP4 \\ 1.08,1.56 \\ 41.6,58.4}
  & \makecell{2$\times$TP4 \\ 1.35,1.29 \\ 50.7,49.3}
  & \makecell{2$\times$TP4 \\ 1.23,1.56 \\ 43.0,57.0}
  & \makecell{2$\times$TP4 \\ 1.2,1.56 \\  43.8,56.2} \\

\bottomrule
\end{tabular}
}
\end{table*}

\appendix
\section*{Appendix}
\section{Placement Configurations}
\autoref{tab:prod-configs} lists the full Tier~1 placement plans used in the
time-varying production-trace experiments reported in
\autoref{subsec:e2e-results}.
It reports, for each load level (67\% and 85\%) and each 5-minute window from
5--10 to 30--35 minutes, the selected prefill/decode configuration for
DistServe and \namebaseline.
Each table entry includes the number and TP degree of instances, the
per-instance GPU frequencies (GHz), and the router load-balancing weights used
during that window.

\end{document}